\documentclass[article,12pt]{elsarticle}
\usepackage{a4wide}
\usepackage{natbib} 			
\usepackage{graphicx}
\usepackage{amsfonts}
\usepackage{amsmath}
\usepackage{booktabs}
\usepackage{epsfig}
\usepackage[colorlinks]{hyperref}
\hypersetup{citecolor=black,linkcolor= black}

\begin{document}
\title{Measuring correlations between non-stationary series with DCCA coefficient}
\author{Ladislav Kristoufek}
\ead{kristouf@utia.cas.cz}
\address{Institute of Information Theory and Automation, Academy of Sciences of the Czech Republic, Pod Vodarenskou Vezi 4, 182 08, Prague 8, Czech Republic\\
Institute of Economic Studies, Faculty of Social Sciences, Charles University, Opletalova 26, 110 00, Prague 1, Czech Republic
}

\begin{abstract}
In this short report, we investigate the ability of the DCCA coefficient to measure correlation level between non-stationary series. Based on a wide Monte Carlo simulation study, we show that the DCCA coefficient can estimate the correlation coefficient accurately regardless the strength of non-stationarity (measured by the fractional differencing parameter $d$). For a comparison, we also report the results for the standard Pearson's correlation coefficient. The DCCA coefficient dominates the Pearson's coefficient for non-stationary series.
\end{abstract}


\journal{Physica A}

\maketitle


\section{Introduction}

Analysis of power-law cross-correlations between various time series has become a popular topic in a broad range of fields -- hydrology \cite{Hajian2010}, seismology \cite{Shadkhoo2009}, traffic \cite{Zebende2009,Xu2010,Zhao2011}, finance \cite{Podobnik2009a,He2011,He2011a,Cao2012}, biometrics \cite{Ursilean2009}, (hydro)meteorology \cite{Vassoler2012,Kang2013}, biology \cite{Xue2012}, DNA sequences \cite{Stan2013}, electricity \cite{Wang2013}, neuroscience \cite{Jun2012}, geophysics \cite{Marinho2013} and others. The power-law (long-term or long-range) cross-correlated processes are standardly described as the ones with power-law decaying cross-correlation function $\rho_{xy}(k)$ with lag $k$ so that $\rho_{xy}(k) \propto k^{-\gamma}$ for $k\rightarrow +\infty$. In contrast, the short-range cross-correlated processes are characteristic by a rapid decay (standardly exponential or faster) of the cross-correlation function. Strength of the power-law cross-correlations is usually represented by the parameter $\lambda$ or the bivariate Hurst exponent $H_{xy}$ which are connected to $\gamma$ as $\lambda=H_{xy}=1-\gamma/2$ \cite{Kristoufek2013}. 

For an estimation of parameters $\lambda$ and $H_{xy}$, several estimators have been proposed -- (multifractal) detrended cross-correlation analysis (DCCA and MF-DXA) \cite{Podobnik2008,Zhou2008}, (multifractal) height cross-correlation analysis (HXA and MF-HXA) \cite{Kristoufek2011}, detrending moving-average cross-correlation analysis (DMCA) \cite{He2011a}, averaged periodogram estimator (APE) \cite{Sela2012} and multifractal cross-correlation analysis based on statistical moments (MFSMXA) \cite{Wang2012}. Several processes mimicking the power-law cross-correlations have been proposed as well \cite{Podobnik2008a,Sela2012,Kristoufek2013a}. Utilizing the ideas of long-range cross-correlations, connected scaling laws and a specific bivariate Hurst exponent estimator DCCA, Zebende \cite{Zebende2011} proposed DCCA cross-correlation coefficient to measure correlation level between non-stationary series. Podobnik \textit{et al.} \cite{Podobnik2011} expanded the coefficient and proposed to use it to test the presence of power-law cross-correlations between series. In this branch of research, the rescaled covariance test has been introduced just recently \cite{Kristoufek2013}. Three ways how to utilize the DCCA coefficient have been further proposed by Balocchi \textit{et al.} \cite{Balocchi2013}, Zebende \textit{et al.} \cite{Zebende2013} and Blythe \cite{Blythe2013}.  Here we focus on the assertion that the DCCA cross-correlation coefficient is able to measure strength and level of correlations between non-stationary series \citep{Zebende2011,Podobnik2011}. 

The paper is organized as follows. Section 2 shortly introduces the DCCA correlation coefficient and the Monte Carlo simulations setting is described in detail. Section 3 brings the results comparing the performance of the DCCA coefficient with respect to changing stationary/non-stationarity. In Section 4, we compare the performance with the standard Pearson's correlation coefficient and we show that the DCCA method strongly outperforms the standard correlation for the non-stationary series.

\section{Methodology}
\subsection{DCCA coefficient}

Detrended cross-correlation coefficient \cite{Zebende2011} combines detrended cross-correlation analysis (DCCA) \cite{Podobnik2008} and detrended fluctuation analysis (DFA) \cite{Peng1993,Peng1994,Kantelhardt2002} to construct a correlation coefficient for detrended series, which might be also asymptotically non-stationary (with the Hurst exponent $H\ge 1$), as
\begin{equation}
\rho_{DCCA}(s)=\frac{F^2_{DCCA}(s)}{F_{DFA,x}(s)F_{DFA,y}(s)}
\label{rho}
\end{equation}
where $F^2_{DCCA}(s)$ is a detrended covariance between partial sums (profiles) $\{X_t\}$ and $\{Y_t\}$ for a window size $s$, and $F^2_{DFA,x}$ and $F^2_{DFA,y}$ are detrended variances of partial sums $\{X_t\}$ and $\{Y_t\}$, respectively, for a window size $s$ \cite{Kantelhardt2002,Podobnik2008}. Specifically for time series $\{x_t\}$, we construct a profile $\{X_t\}$ as $X_t=\sum_{i=1}^t{(x_i-\bar{x})}$ which is divided into overlapping boxes of length (scale) $s$. In each box between $j$ and $j+s-1$, the linear fit of a time trend is constructed so that we get $\widehat{X_{k,j}}$ for $j\le k \le j+s-1$. Detrended variance is then defined as
\begin{equation}
f_{DFA,x}^2(s,j)=\frac{\sum_{k=j}^{j+s-1}{(X_k-\widehat{X_{k,j}})^2}}{s-1}.
\label{DFA1}
\end{equation}
Detrended variance is then averaged over all boxes of length $s$ to obtain the fluctuation $F_{DFA,x}^2(s)$
\begin{equation}
F_{DFA,x}^2(s)=\frac{\sum_{j=1}^{T-s+1}{f_{DFA,x}^2(s,j)}}{T-s}.
\label{DFA2}
\end{equation}
For two time series $\{x_t\}$ and $\{y_t\}$, we are interested in the detrended covariance of profiles
\begin{equation}
f_{DCCA}^2(s,j)=\frac{\sum_{k=j}^{j+s-1}{(X_k-\widehat{X_{k,j}})(Y_k-\widehat{Y_{k,j}})}}{s-1}
\label{DCCA1}
\end{equation}
which is again averaged over all boxes\footnote{For computational efficiency, we use non-overlapping boxes of size $s$. In the cases when $T/s$ is not an integer, we average the fluctuations of boxes split both from the beginning and the end of the series so that we obtain $2\lfloor T/s\rfloor$ boxes.} with scale $s$ to obtain
\begin{equation}
F_{DCCA}^2(s)=\frac{\sum_{j=1}^{T-s+1}{f_{DCCA}^2(s,j)}}{T-s}.
\label{DCCA2}
\end{equation}
These fluctuations are then inputted into Eq. \ref{rho}. Podobnik \textit{et al.} \cite{Podobnik2011} show that $-1 \le \rho_{DCCA}(s)\le 1$ so that the DCCA coefficient can be interpreted as a standard correlation coefficient with $\rho_{DCCA}(s)=-1$ for perfectly anti-correlated series, $\rho_{DCCA}(s)=1$ for perfectly correlated series and $\rho_{DCCA}(s)=0$ for uncorrelated processes.

\subsection{Simulations setting}

We are interested in the ability of the DCCA coefficient to measure correlation between non-stationary series. To do so, we simulate a wide range of processes with varying correlation strength and varying level of (non-)stationarity. Two ARFIMA(0,$d$,0) processes with correlated innovations are used for this matter:
\begin{gather}
\label{eq:ARFIMA1}
x_t=\sum_{n=0}^{\infty}{a_n(d_1)\varepsilon_{t-n}}\\
y_t=\sum_{n=0}^{\infty}{a_n(d_2)\nu_{t-n}}
\end{gather}
where $a_n(d)=\frac{\Gamma(n+d)}{\Gamma(n+1)\Gamma(d)}$, $\langle \varepsilon_t \rangle=\langle \nu_t \rangle=0$, $\langle \varepsilon^2_t \rangle=\langle \nu^2_t \rangle=1$ and $\langle \varepsilon_t\nu_t \rangle=\rho_{\varepsilon\nu}$. For our purposes, we vary the parameters $d_1$, $d_2$, $\rho_{\varepsilon\nu}$, $s$ and $T$ to see how the method fares. We use $d_1=d_2=d$ for simplicity and $\rho_{\varepsilon\nu}$ ranges between -0.9 and 0.9 with a step of 0.1. Two time series lengths -- $T=1000$ and $T=5000$ -- are analyzed to see whether the performance of the DCCA coefficient changes for different series lengths. The last parameter we vary is the scale $s$ where we utilize four different levels -- $s=\frac{T}{100},\frac{T}{50},\frac{T}{10},\frac{T}{5}$.

\begin{figure}[!htbp]
\begin{center}
\begin{tabular}{ccc}
\includegraphics[width=45mm]{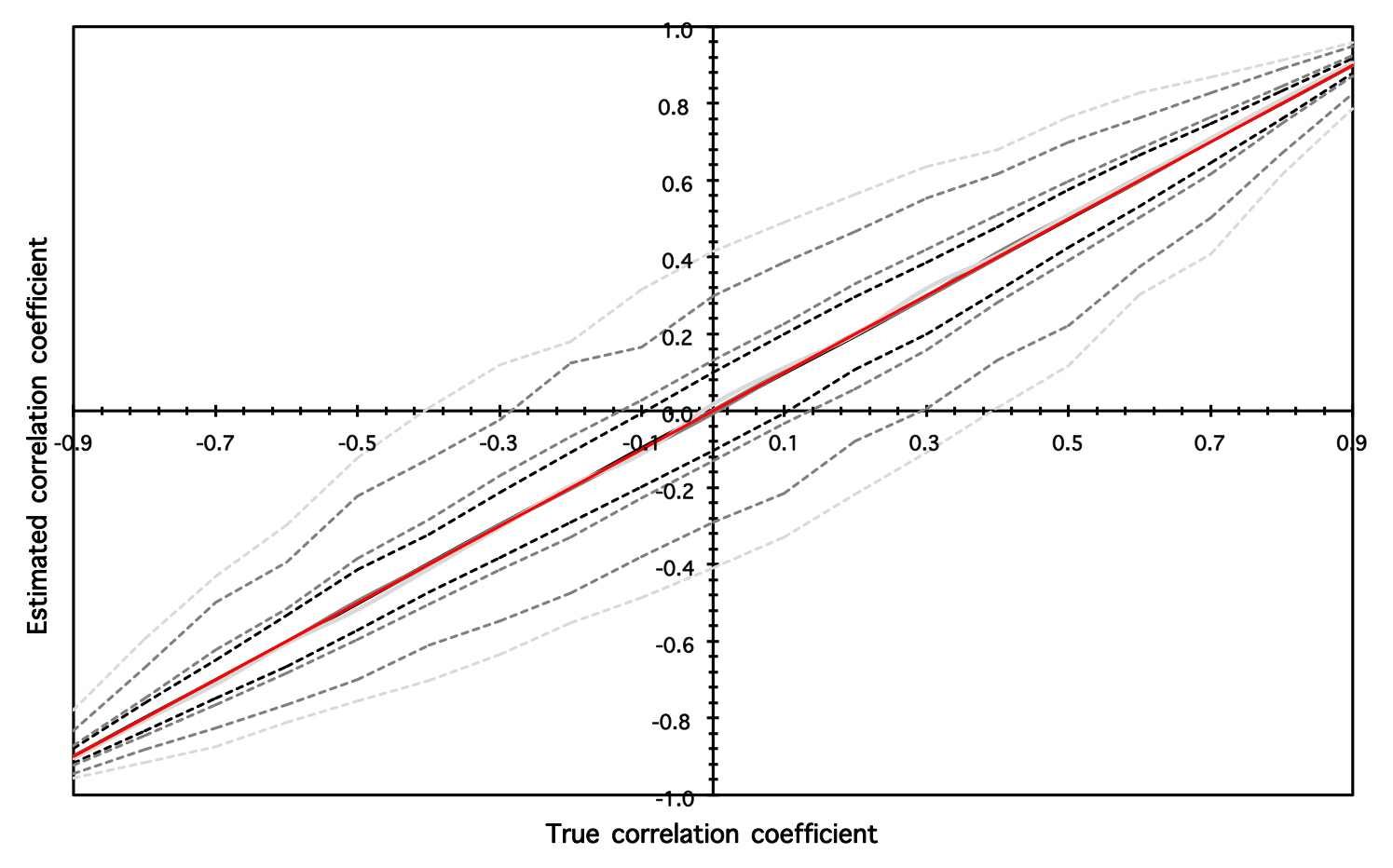}&\includegraphics[width=45mm]{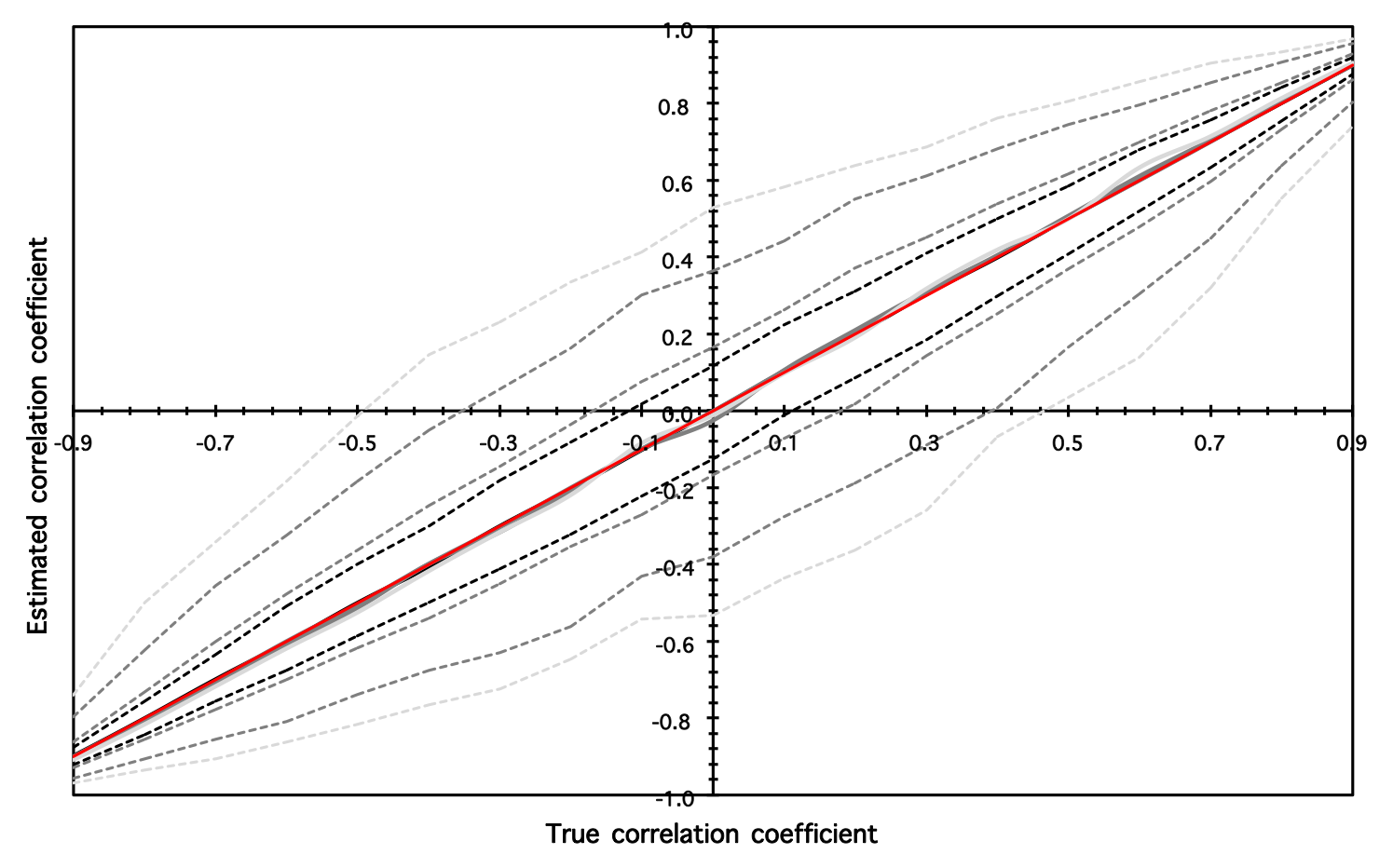}&\includegraphics[width=45mm]{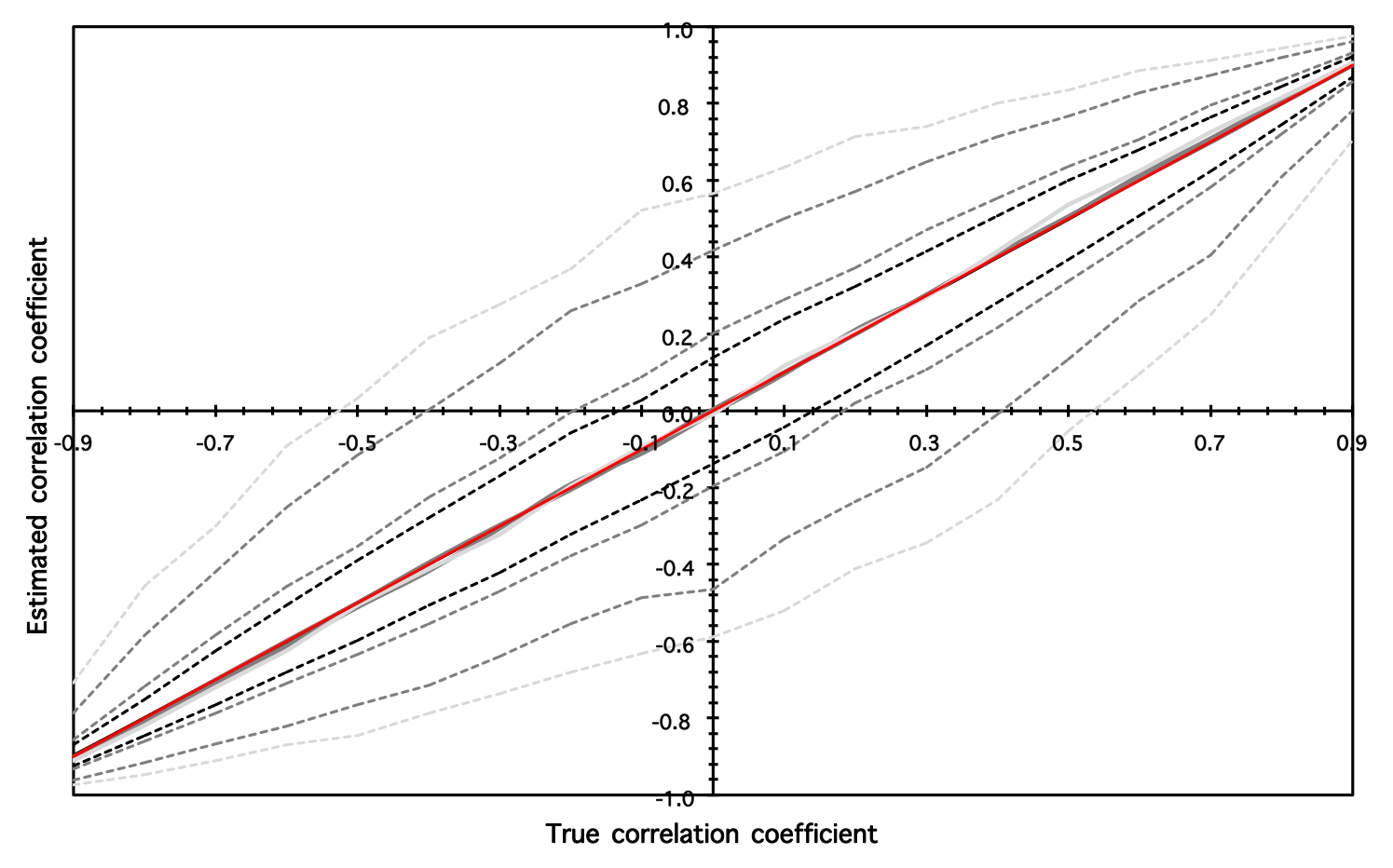}\\
\includegraphics[width=45mm]{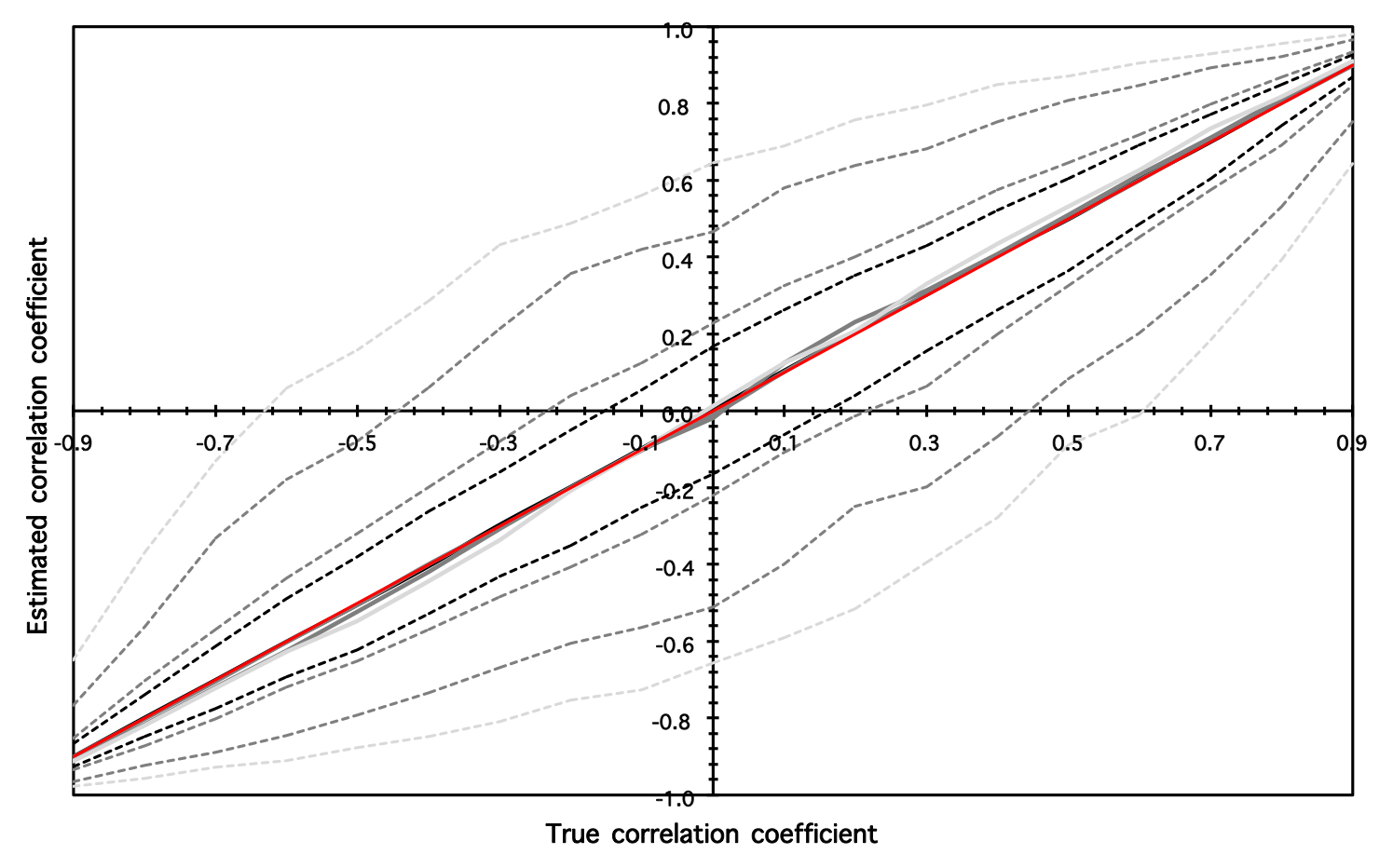}&\includegraphics[width=45mm]{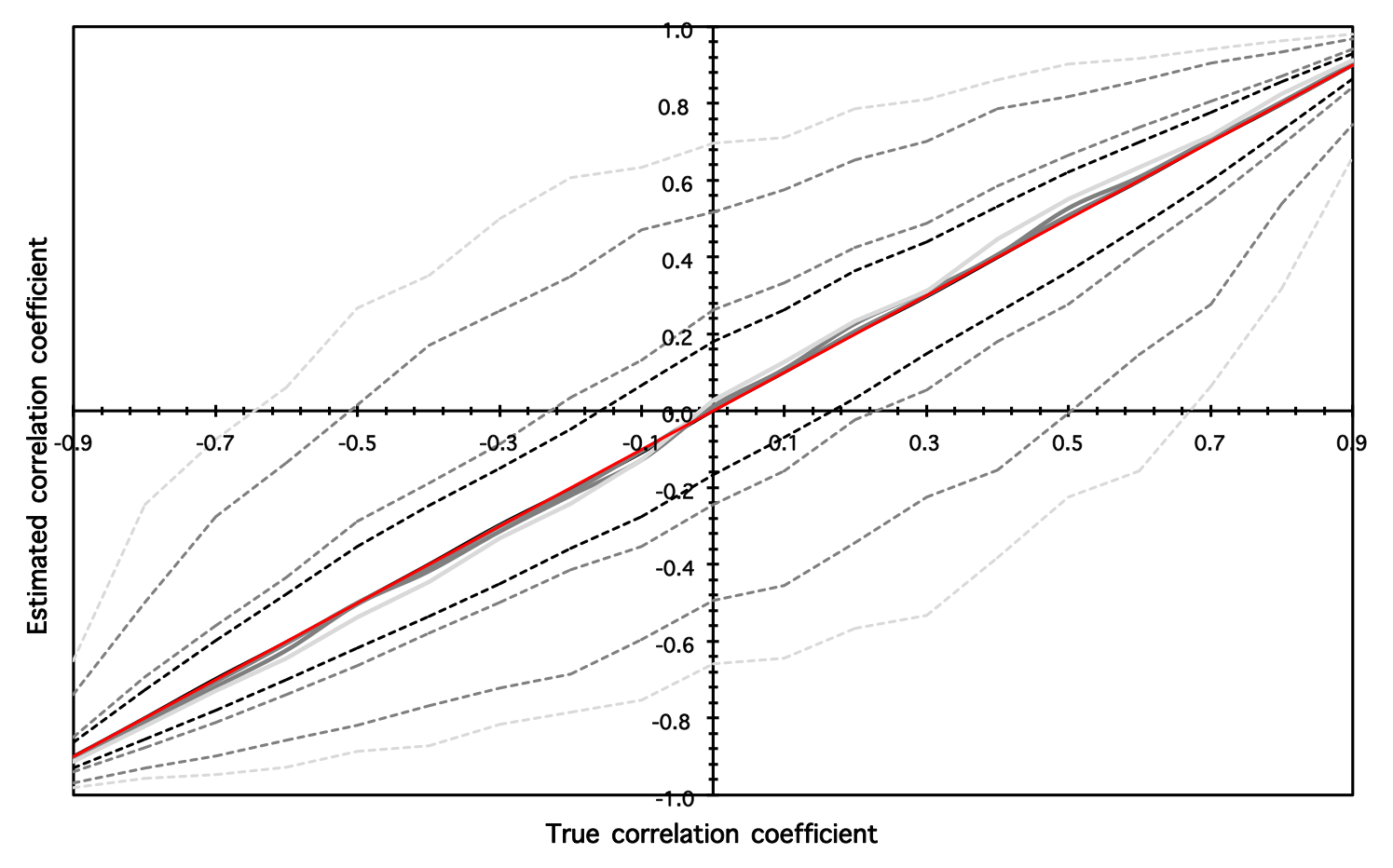}&\includegraphics[width=45mm]{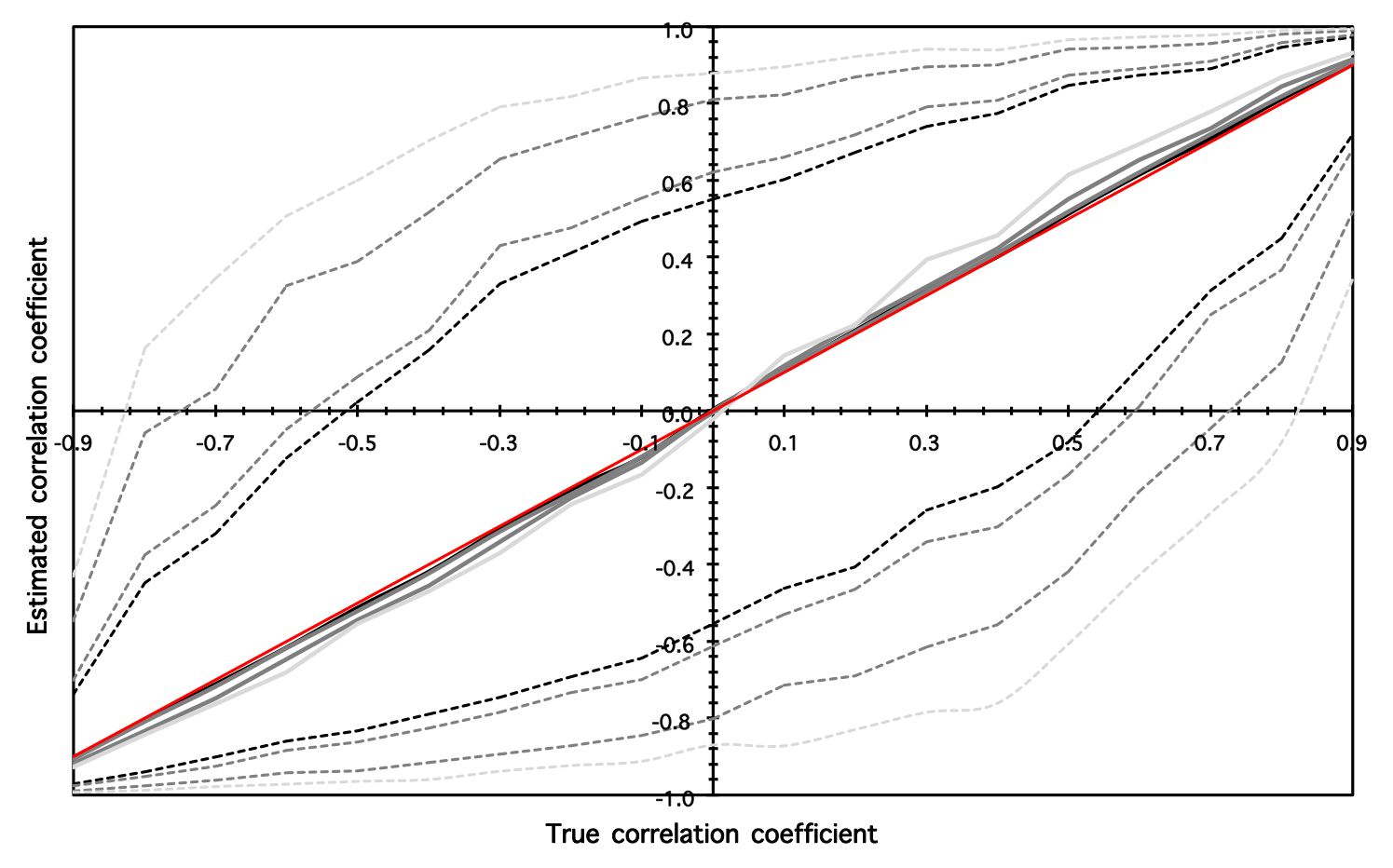}\\
\end{tabular}
\caption{\textbf{Estimated DCCA correlation coefficients for different fractional integration parameters $d$ I.} \footnotesize{Results for the time series of length $T=1000$ are shown here. Separate figures represent different parameters $d$ -- $d=0.1$ (top left), $d=0.4$ (top center), $d=0.6$ (top right), $d=0.9$ (bottom left), $d=1.1$ (bottom center), $d=1.4$ (bottom right). Red lines represent the true value of $\rho_{\varepsilon\nu}$. The solid lines of shades of grey (mostly overlapping with the red line) represent the median values of 1000 simulations for the given parameter setting. The dashed lines represent the 95\% confidence intervals (the 2.5th and the 97.5th quantiles of the simulations). Different shades of grey stand for different values of $s$ going from the lowest scales ($s=T/100$, the darkest shade) to the highest scales ($s=T/5$, the lightest shade).}\label{fig1}}
\end{center}
\end{figure}

In an ideal case, the estimated correlation coefficients should be equal to $\rho_{\varepsilon\nu}$ regardless the other parameters and mainly the parameter $d$. Fractional differencing parameter $d$ separates the time series between stationary and non-stationary, $d=0.5$ (which is parallel to $H=1$ in the long-term memory setting) being the separating point. However, the separation is more detailed. For $d<0.5$, we have stationary processes. For $0.5 \le d < 1$, the processes are non-stationary but mean-reverting. And for $d\ge 1$, we obtain non-stationary non-mean-reverting (explosive) processes. In order to show whether the DCCA coefficient is able to capture the correlation between processes with different levels of (non-)stationarity, we study the cases when $d=0.1,0.4,0.6,0.9,1.1,1.4$, i.e. two stationary, non-stationary mean-reverting, and non-stationary non-mean-reverting processes each.

For each combination of parameters $d$, $\rho_{\varepsilon\nu}$, $s$ and $T$, we simulate 1000 repetitions. Each setting is then analyzed with respect to the 2.5th, the 50th and the 97.5th quantiles, i.e. the 95\% confidence intervals and the median, and the standard deviation of the estimated DCCA coefficient. For the comparison purposes and to show a potential power of the DCCA coefficient, we also discuss the results of the standard Pearson's correlation coefficient. 

\section{Results}

We are primarily interested in the ability of the DCCA coefficient to estimate a correct correlation between two series regardless the potential non-stationarity. For that matter, we perform a wide Monte Carlo simulation study where the memory parameter $d$, correlation coefficient $\rho_{\varepsilon\nu}$, scale $s$ and time series length $T$ vary, and we are interested in the influence of these parameters on the performance of the coefficient $\rho_{DCCA}(s)$. For a better comparison of the results and for easier drawing of conclusion, we present the results in a form of charts.

\begin{figure}[!htbp]
\begin{center}
\begin{tabular}{ccc}
\includegraphics[width=45mm]{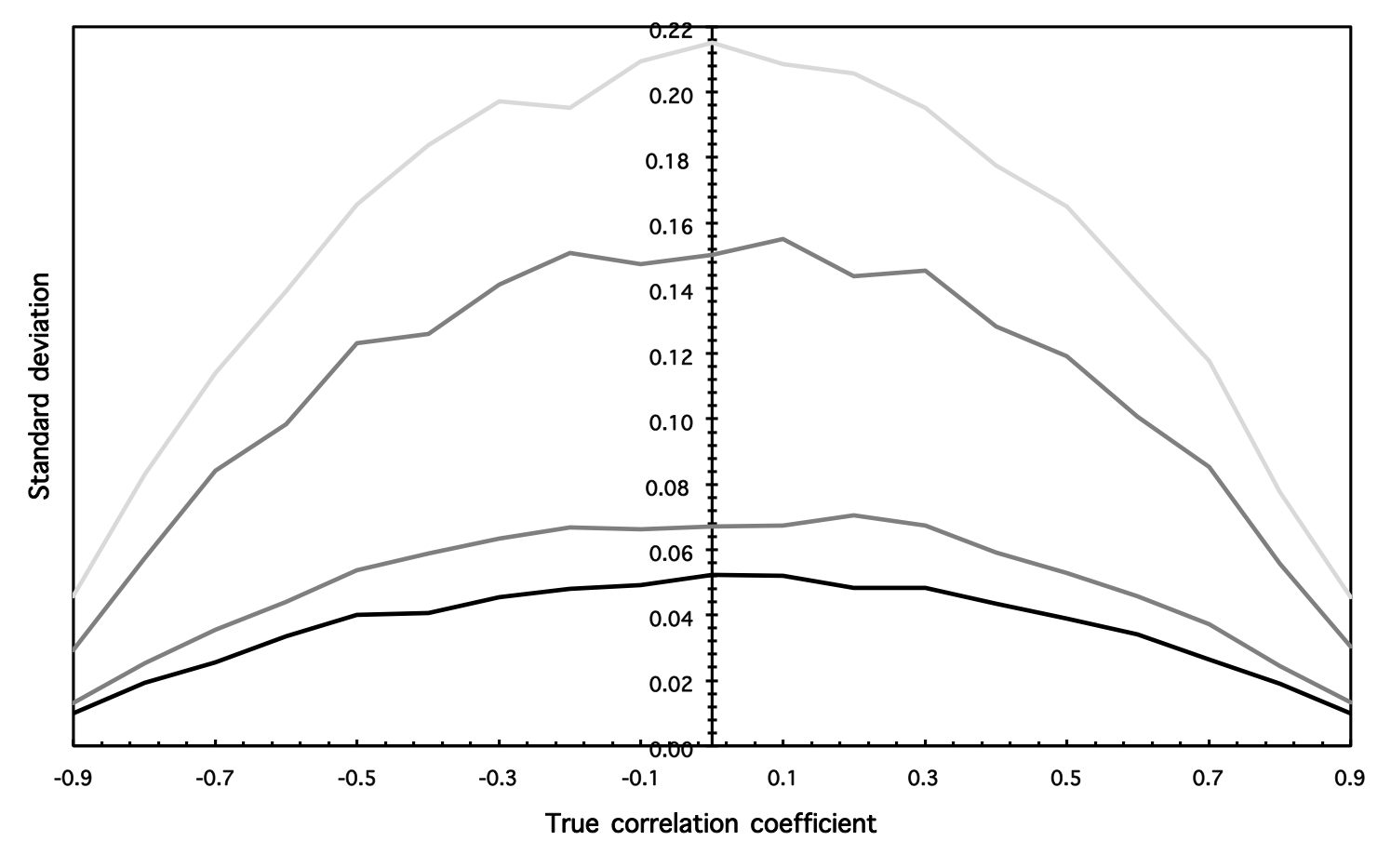}&\includegraphics[width=45mm]{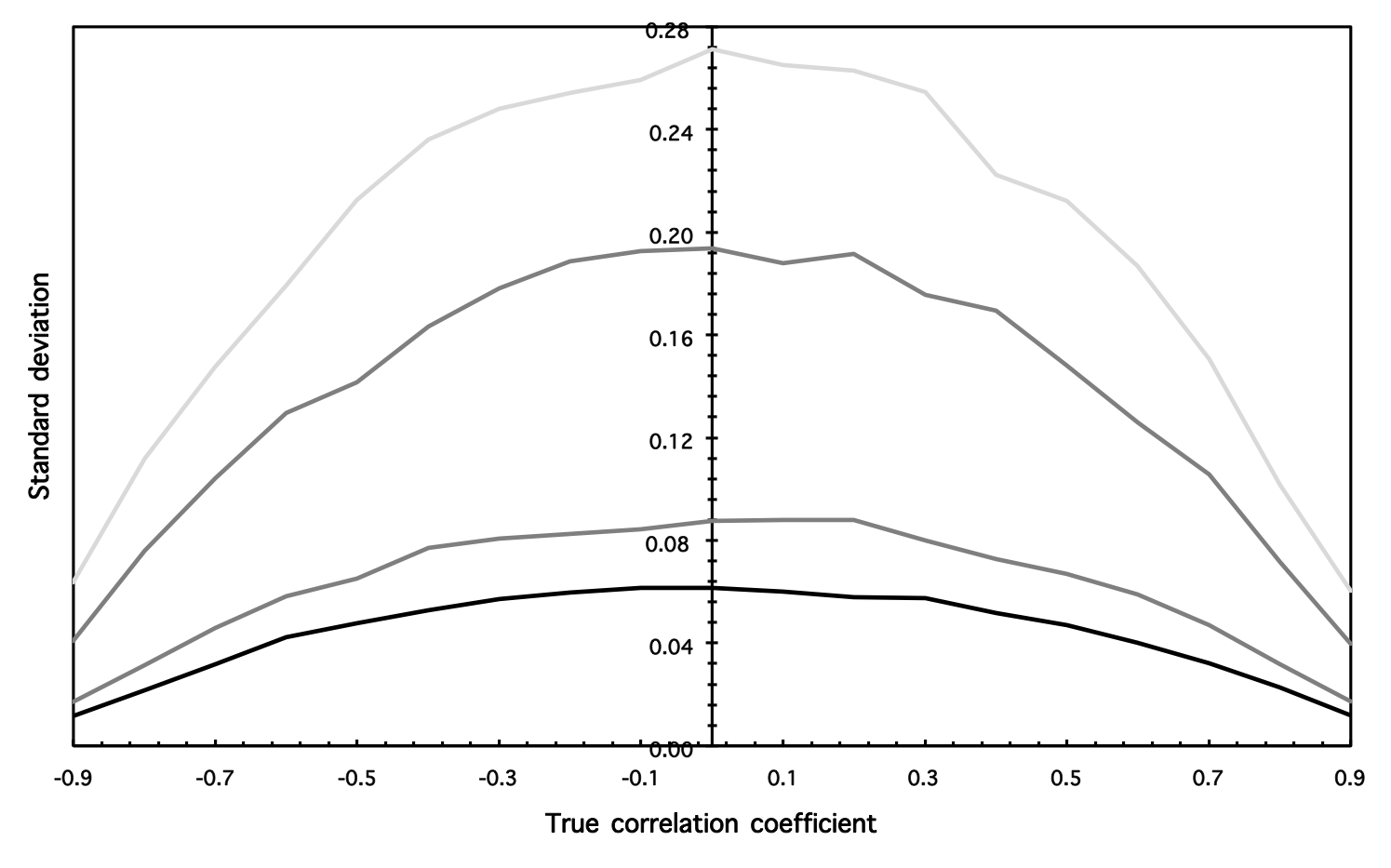}&\includegraphics[width=45mm]{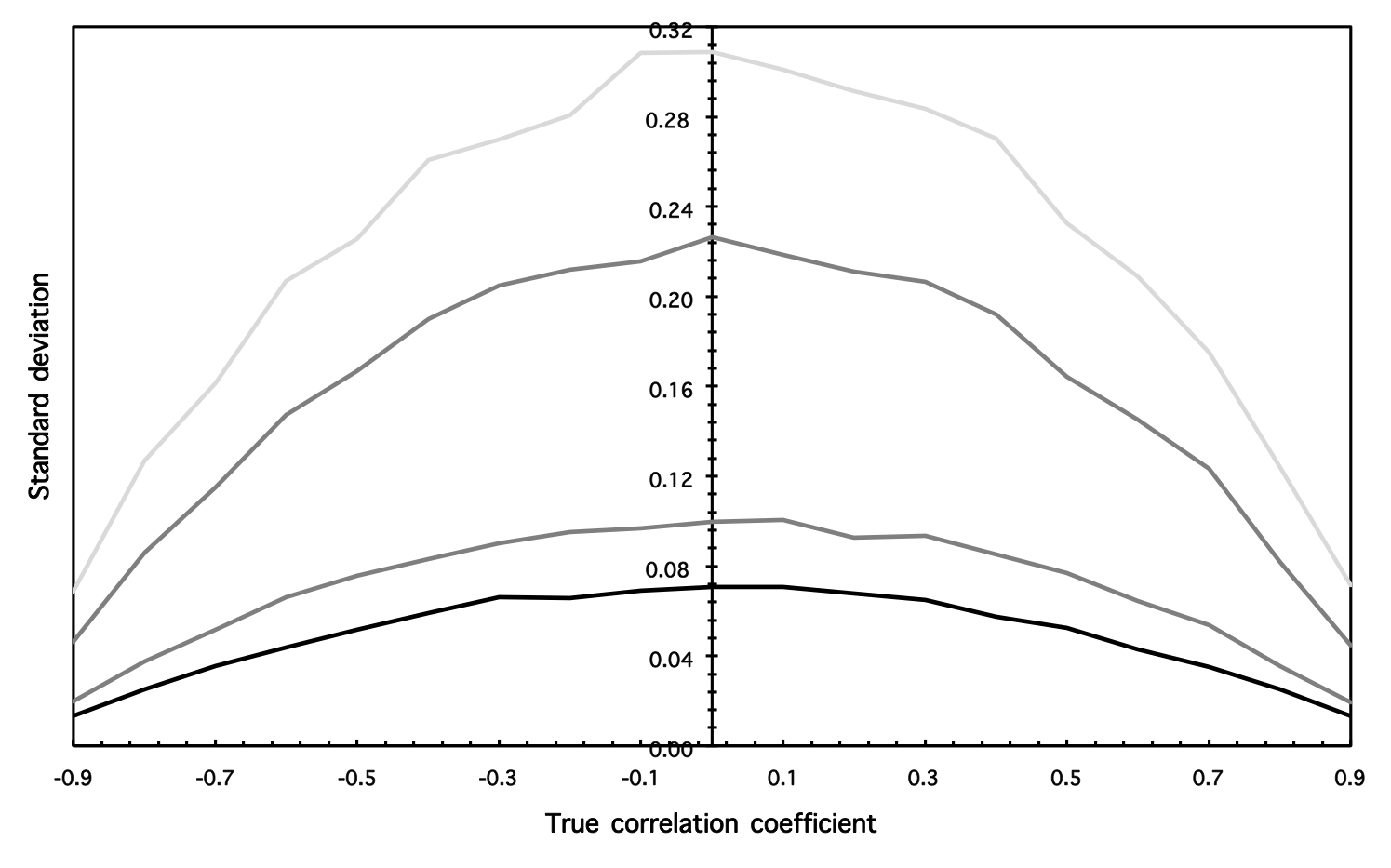}\\
\includegraphics[width=45mm]{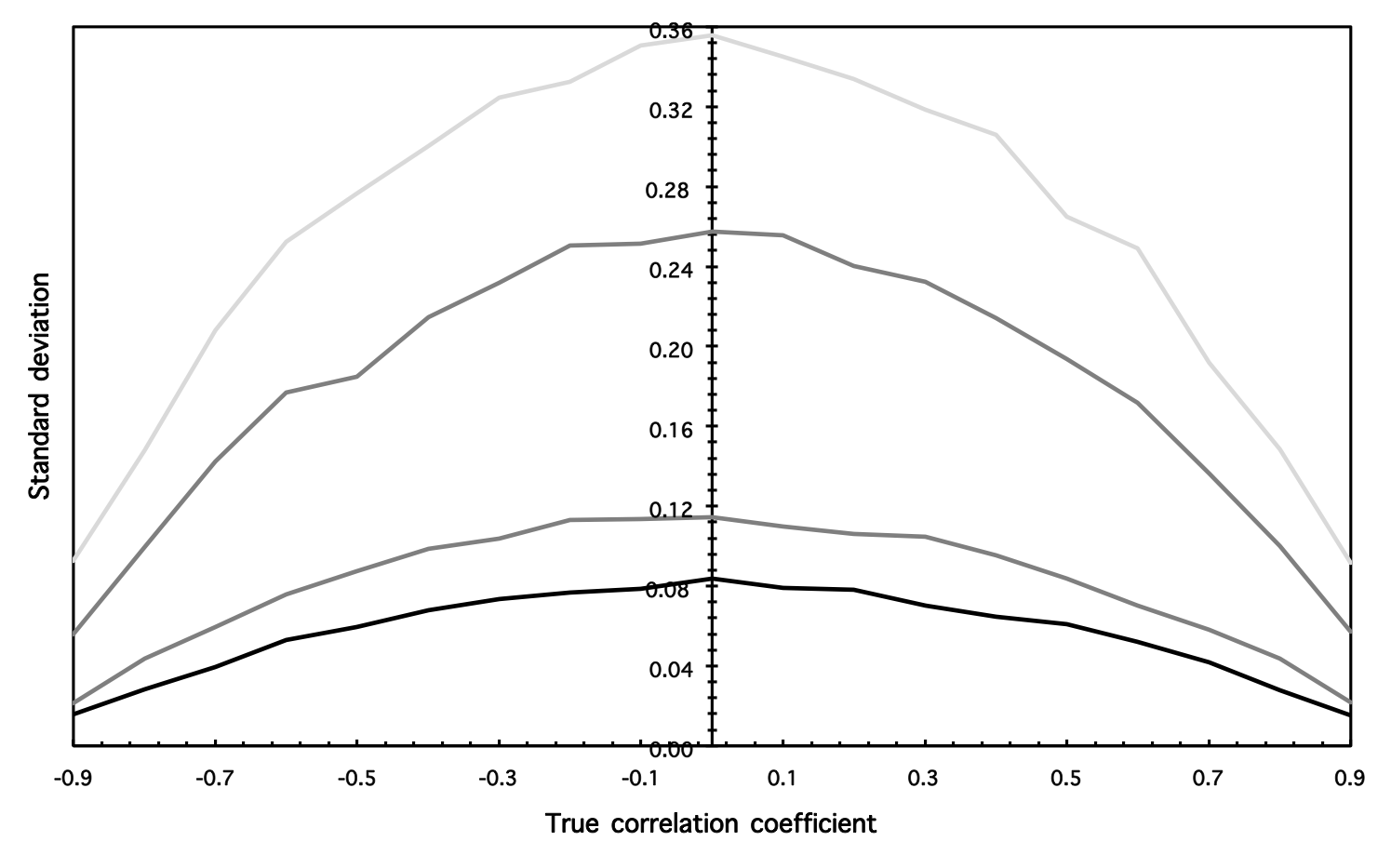}&\includegraphics[width=45mm]{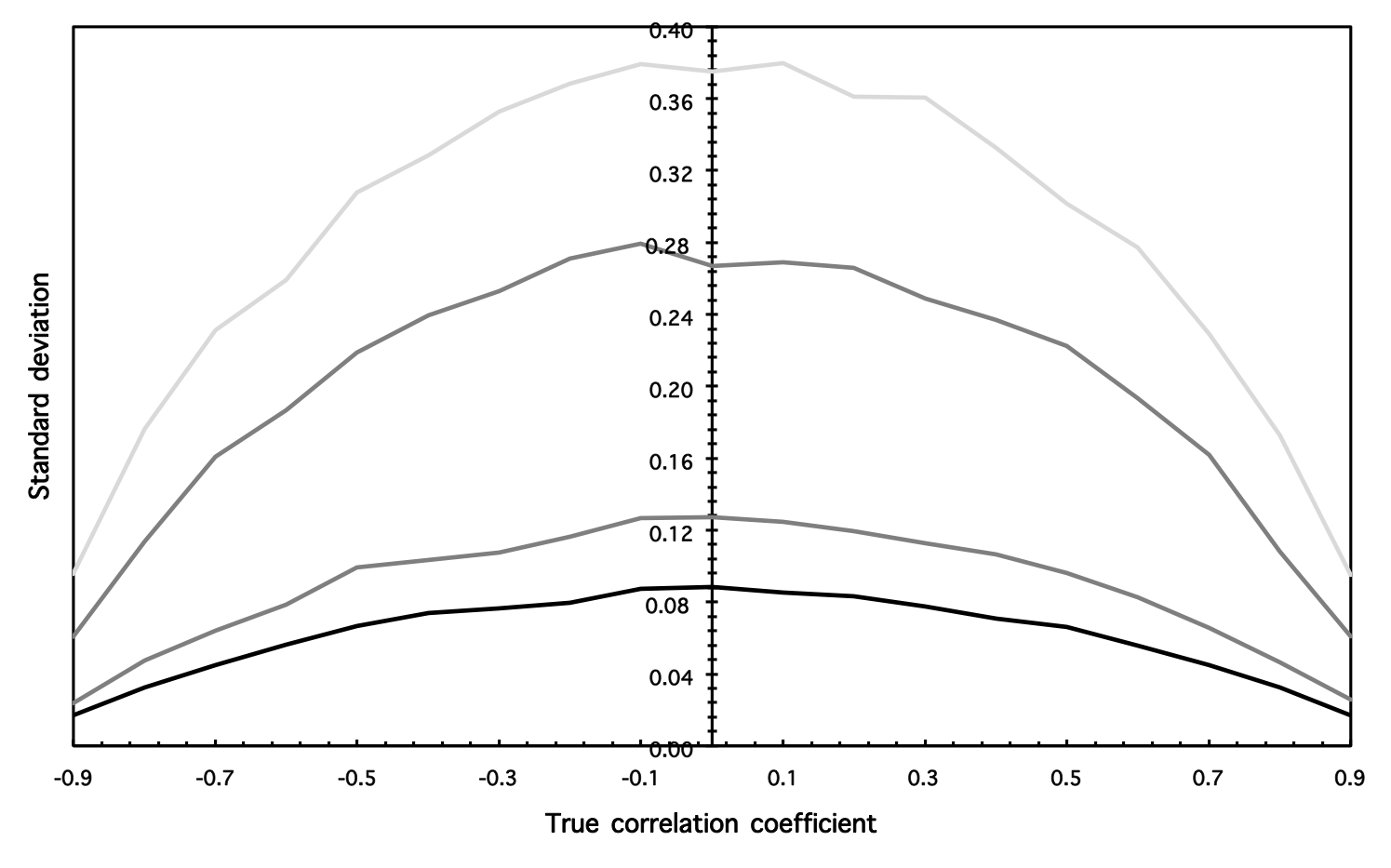}&\includegraphics[width=45mm]{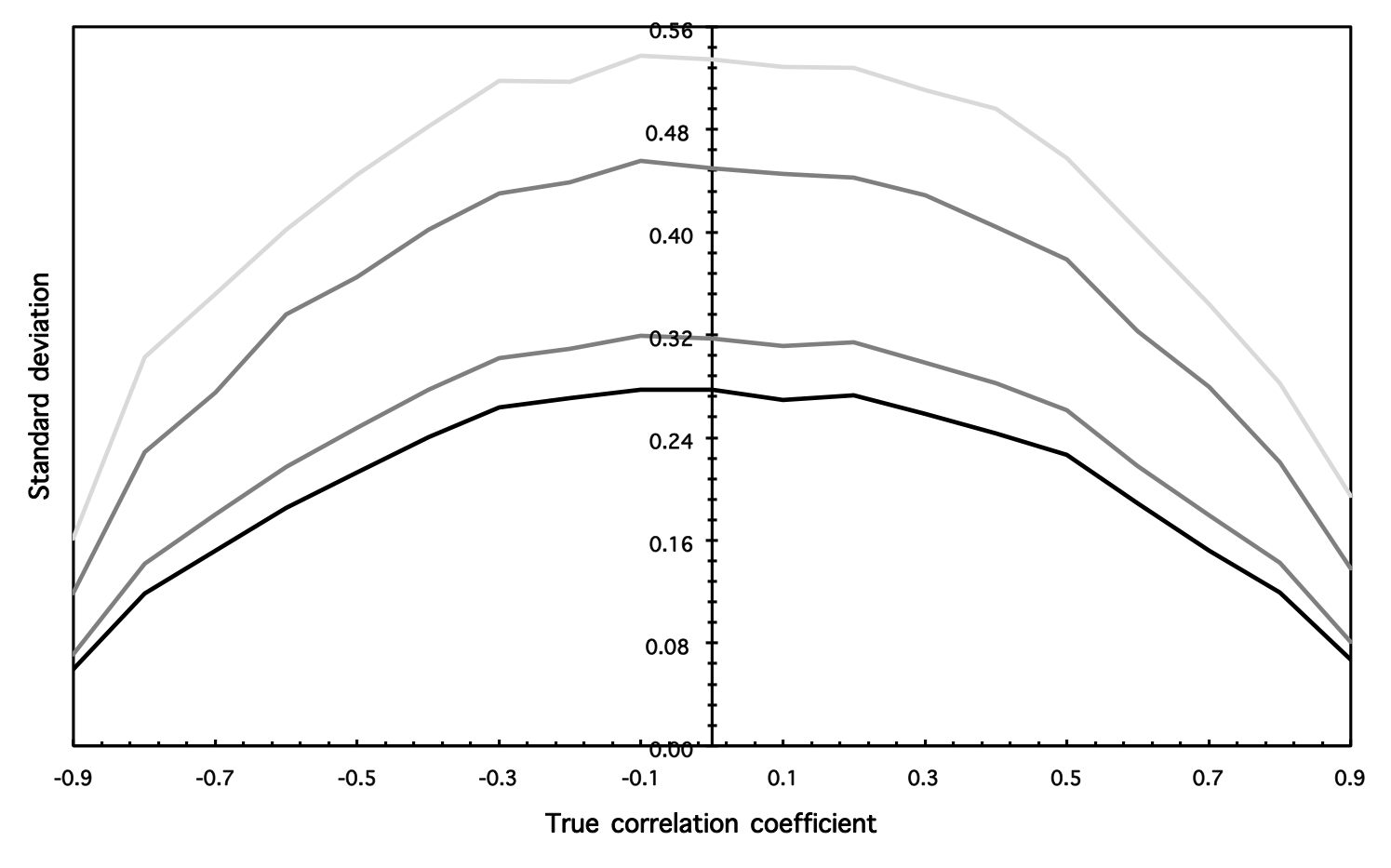}\\
\end{tabular}
\caption{\textbf{Standard deviations of DCCA correlation coefficients for different fractional integration parameters $d$ I.} \footnotesize{Results for the time series of length $T=1000$ are shown here. Separate figures represent different parameters $d$ -- $d=0.1$ (top left), $d=0.4$ (top center), $d=0.6$ (top right), $d=0.9$ (bottom left), $d=1.1$ (bottom center), $d=1.4$ (bottom right). Solid lines represent the standard deviation of 1000 simulations for given parameter setting. Different shades of grey stand for different values of $s$ going from the lowest scales ($s=T/100$, the darkest shade) to the highest scales ($s=T/5$, the lightest shade).}\label{fig2}}
\end{center}
\end{figure}

\begin{figure}[!htbp]
\begin{center}
\begin{tabular}{ccc}
\includegraphics[width=45mm]{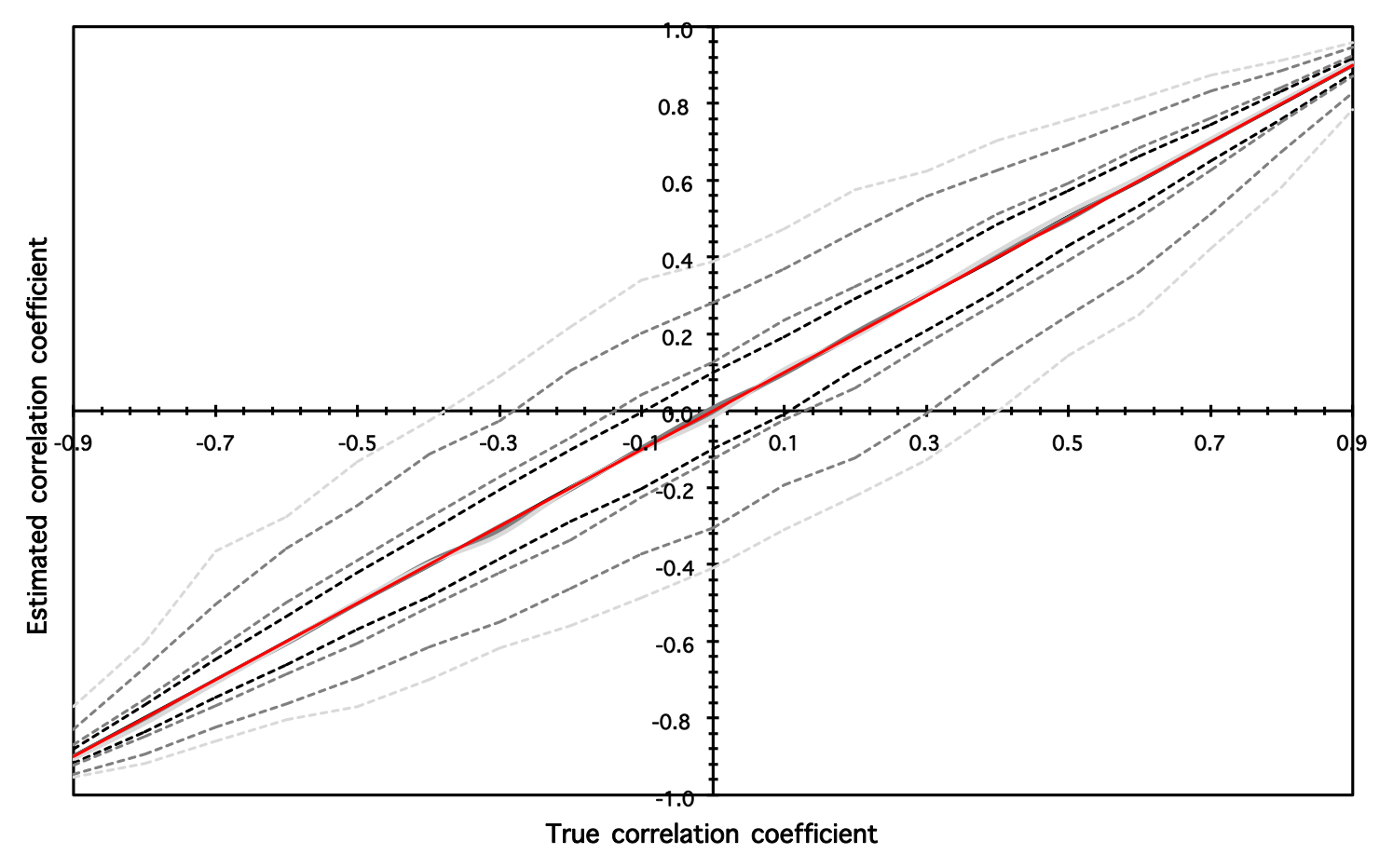}&\includegraphics[width=45mm]{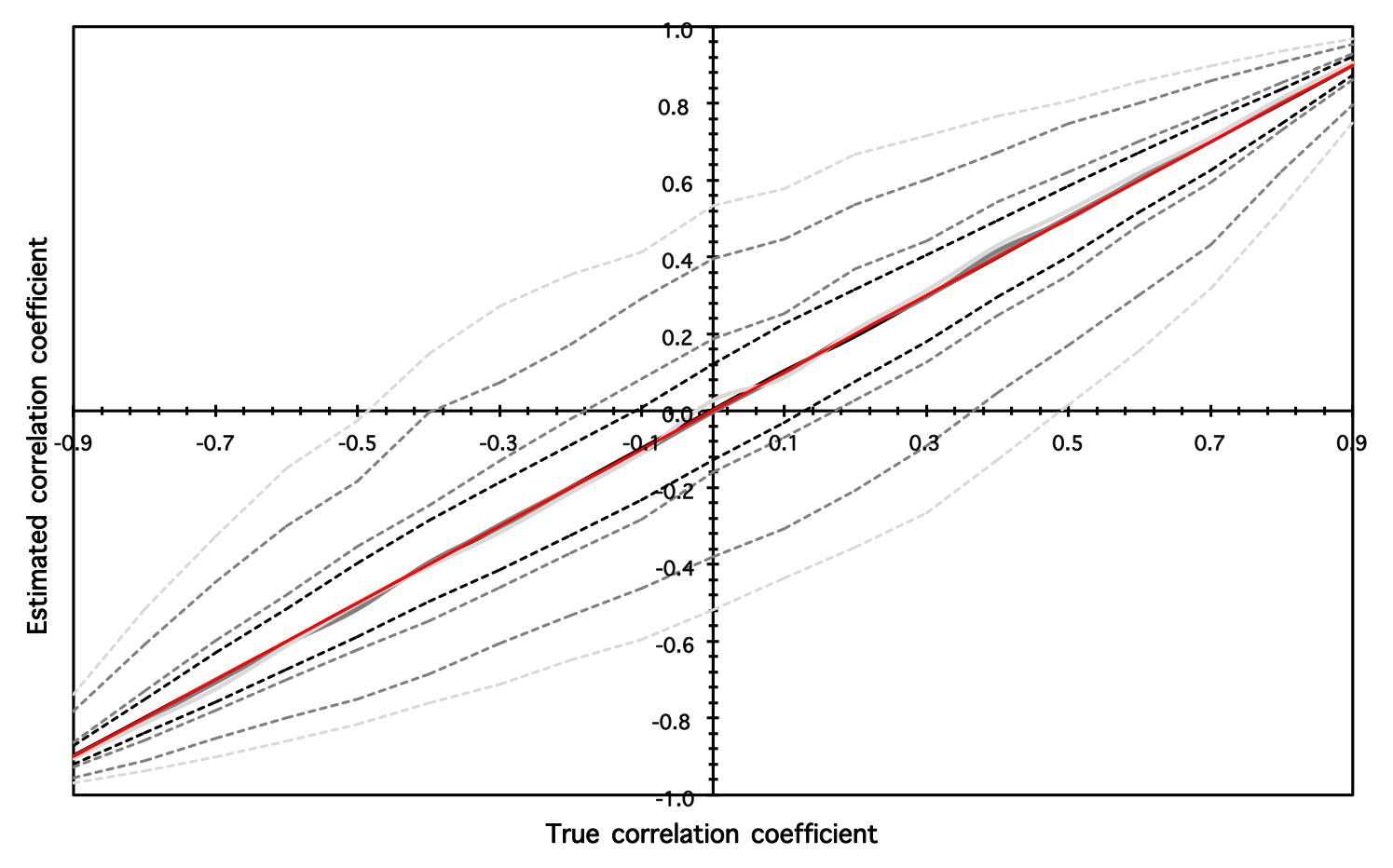}&\includegraphics[width=45mm]{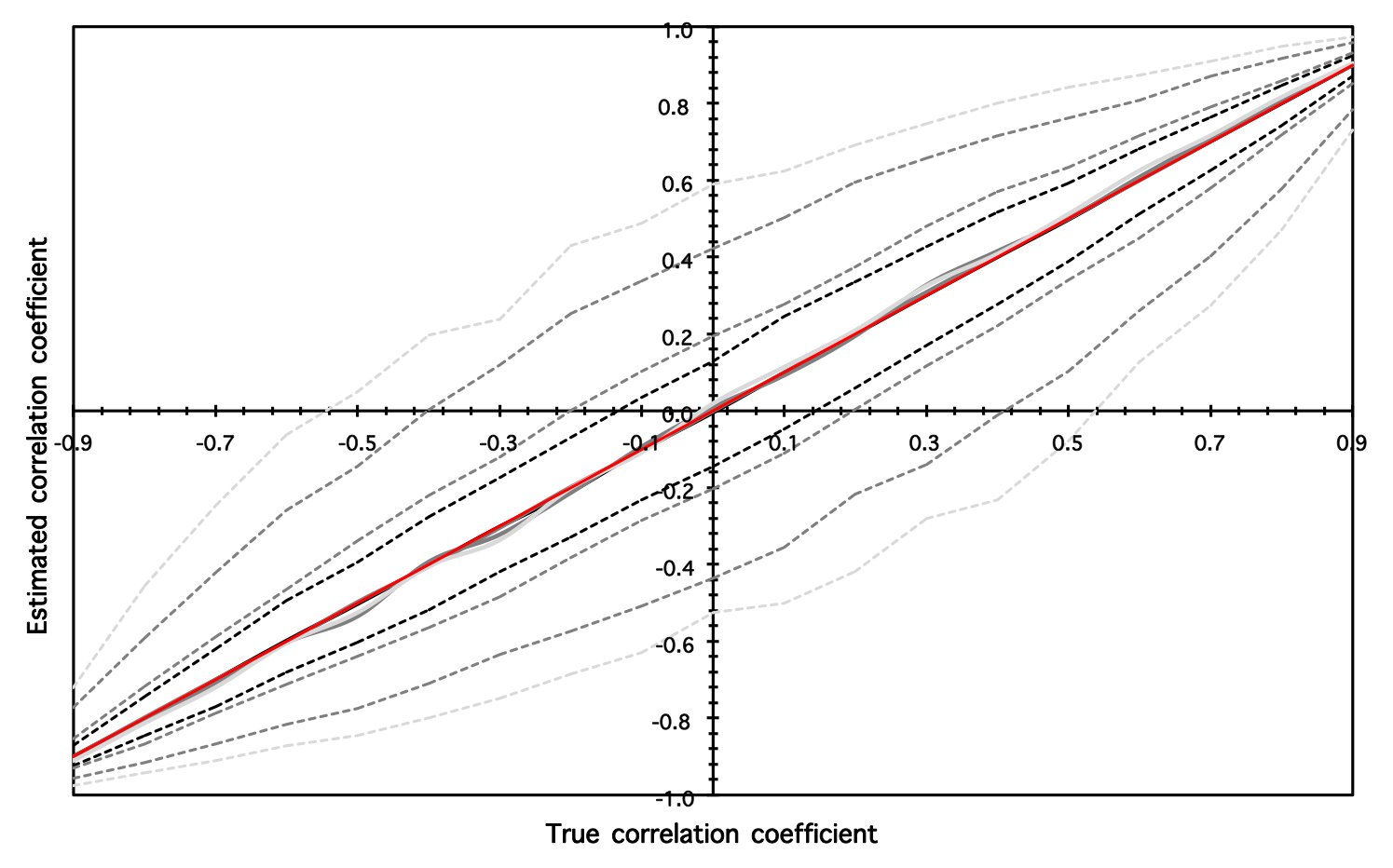}\\
\includegraphics[width=45mm]{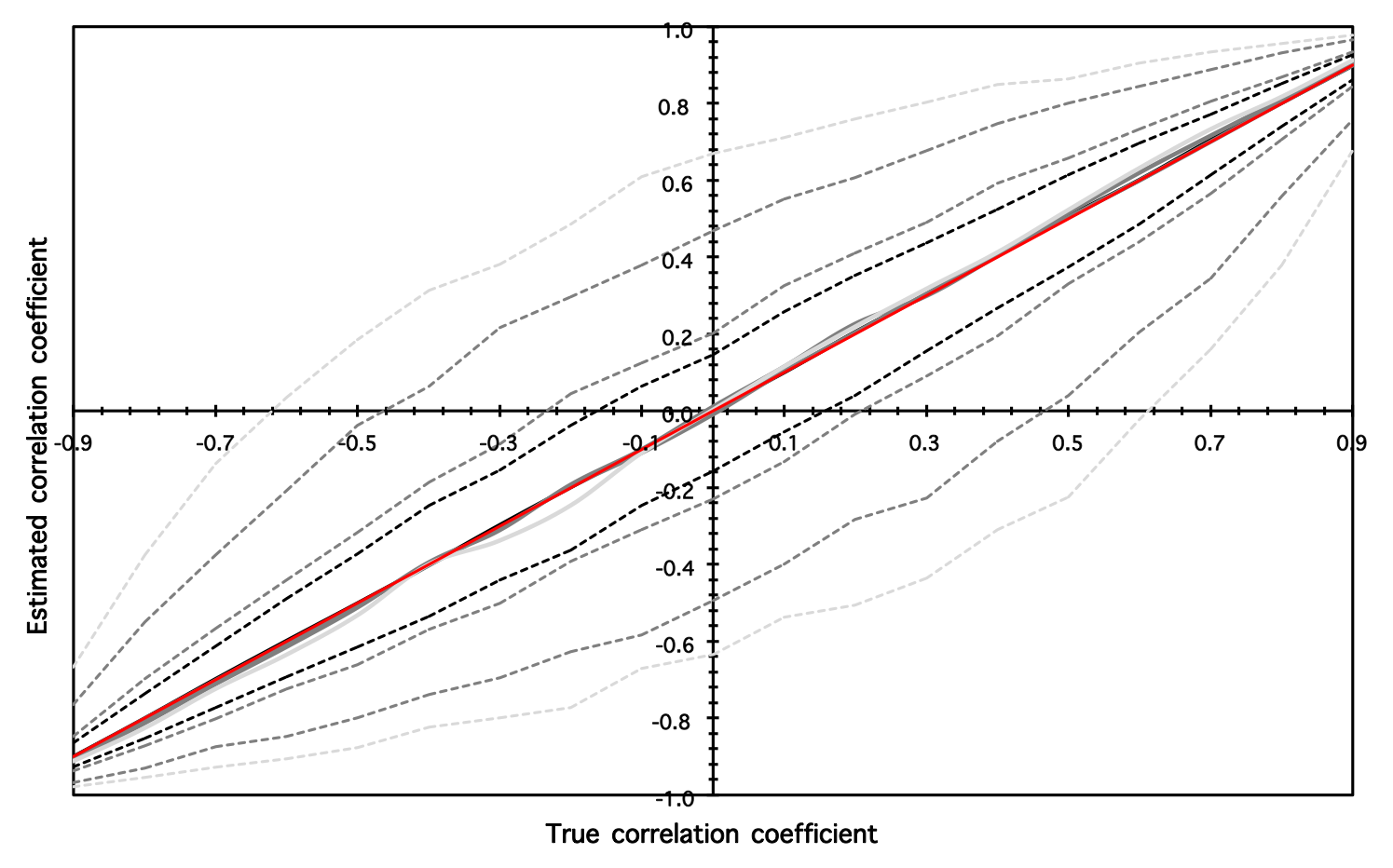}&\includegraphics[width=45mm]{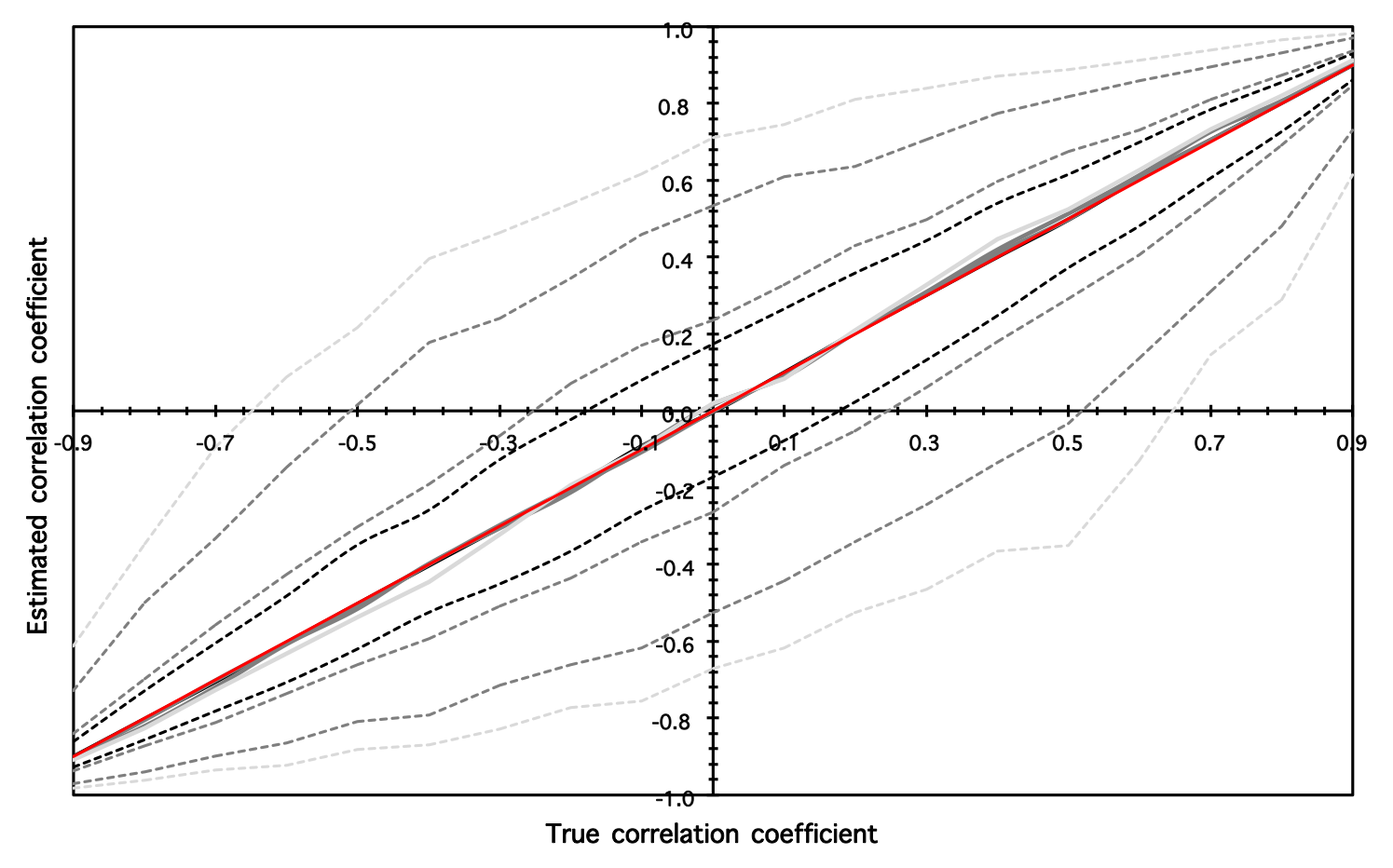}&\includegraphics[width=45mm]{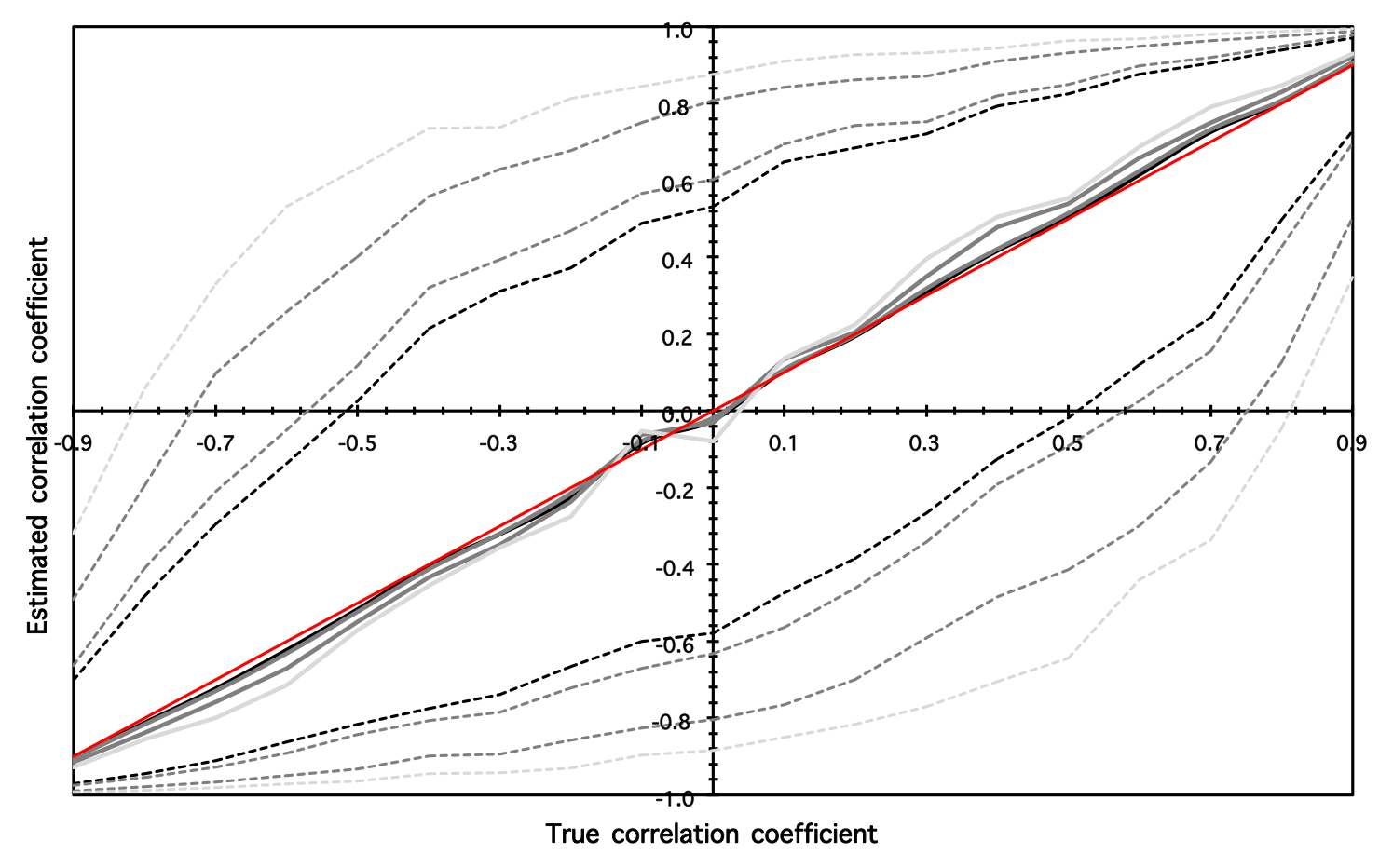}\\
\end{tabular}
\caption{\textbf{Estimated DCCA correlation coefficients for different fractional integration parameters $d$ II.} \footnotesize{Results for the time series of length $T=5000$ are shown here. Notation of Fig. \ref{fig1} is used.}\label{fig3}}
\end{center}
\end{figure}

\begin{figure}[!htbp]
\begin{center}
\begin{tabular}{ccc}
\includegraphics[width=45mm]{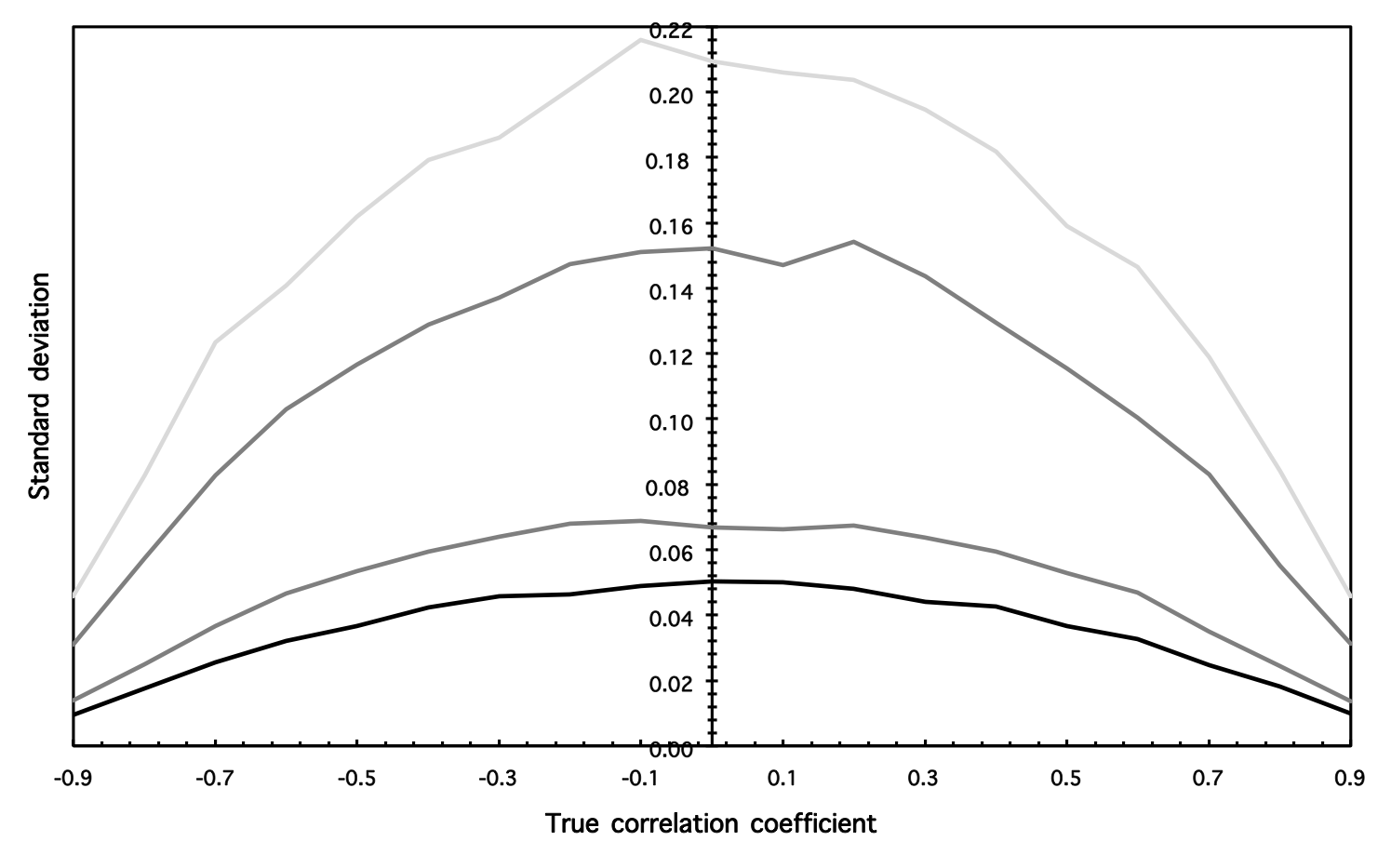}&\includegraphics[width=45mm]{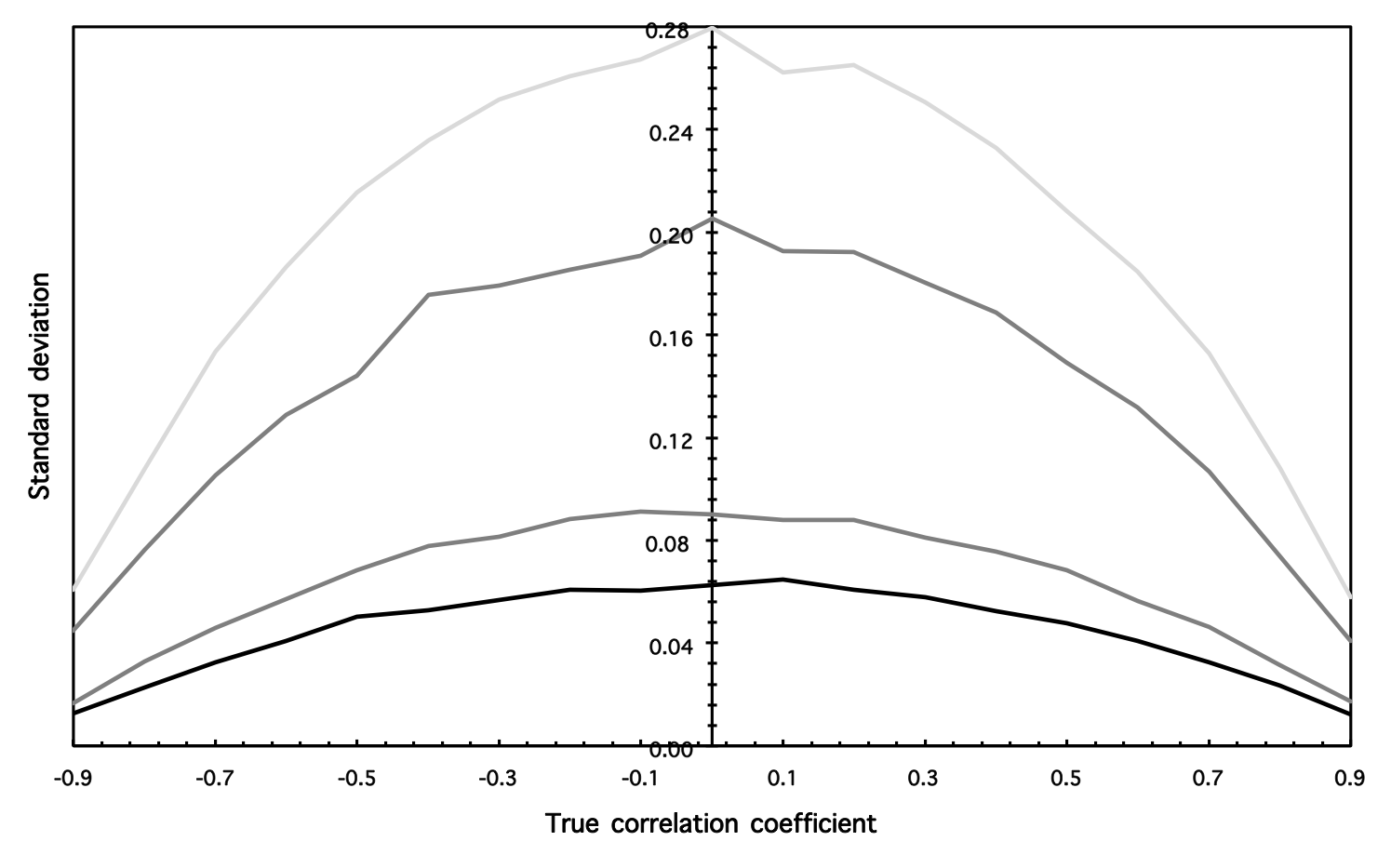}&\includegraphics[width=45mm]{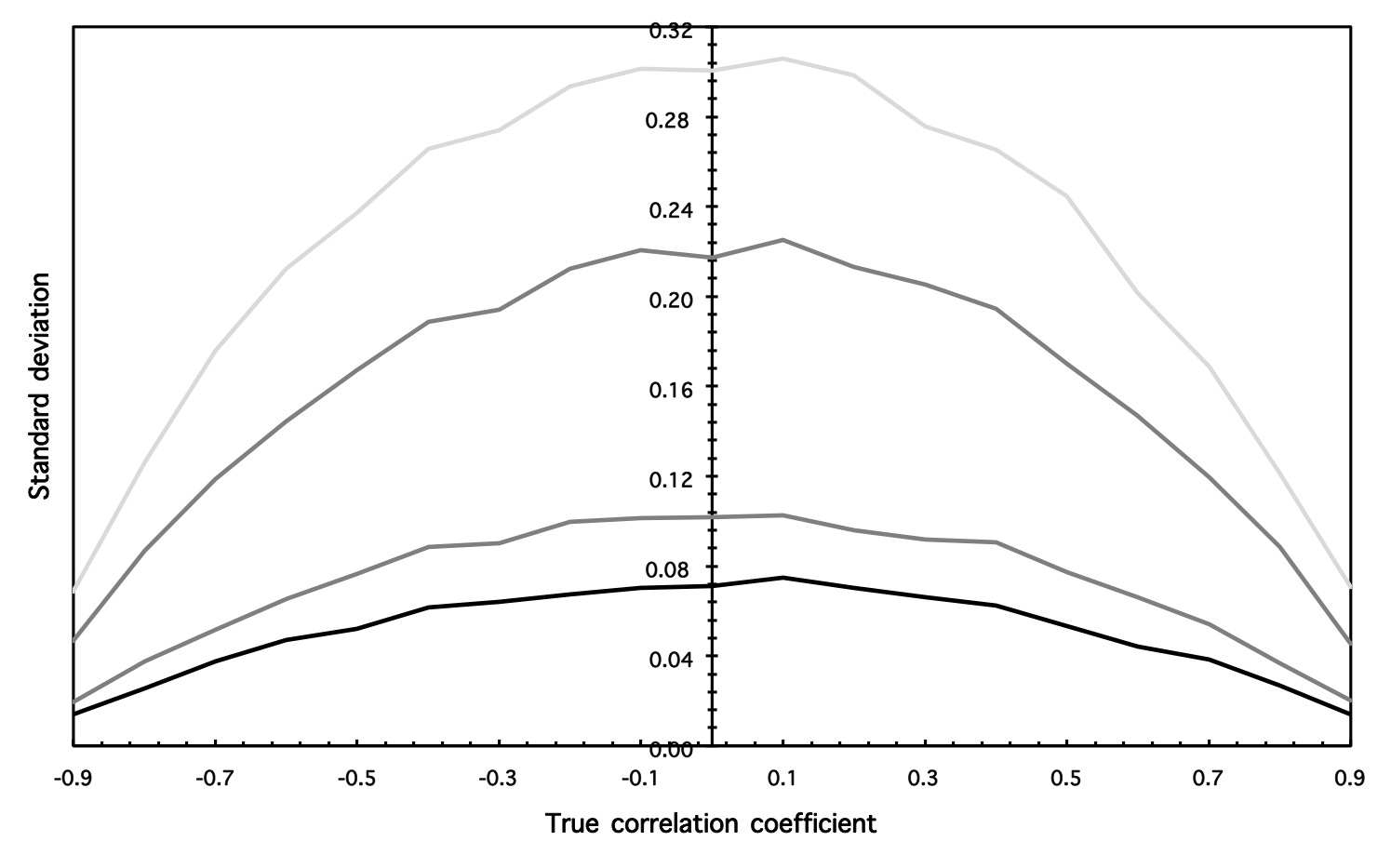}\\
\includegraphics[width=45mm]{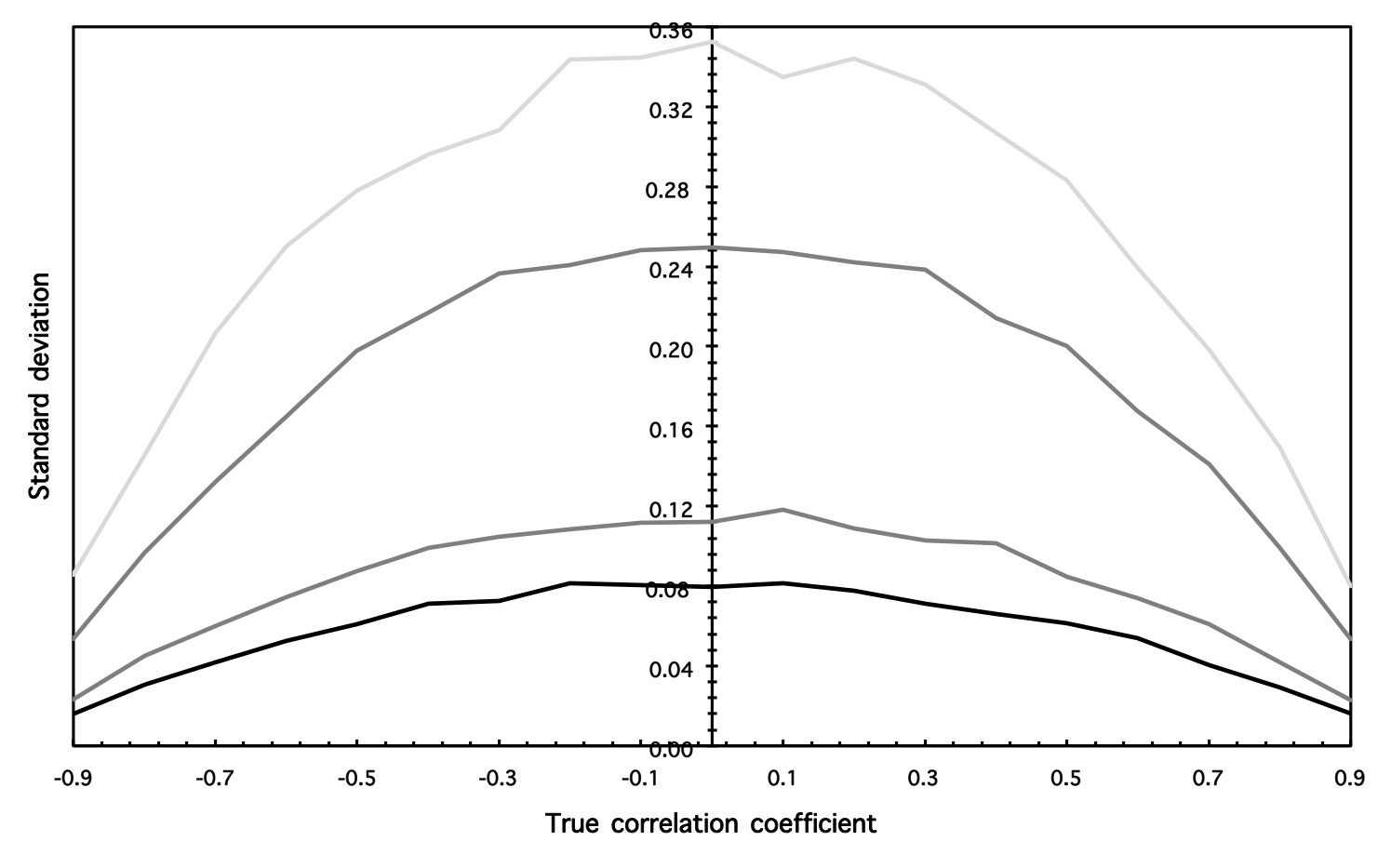}&\includegraphics[width=45mm]{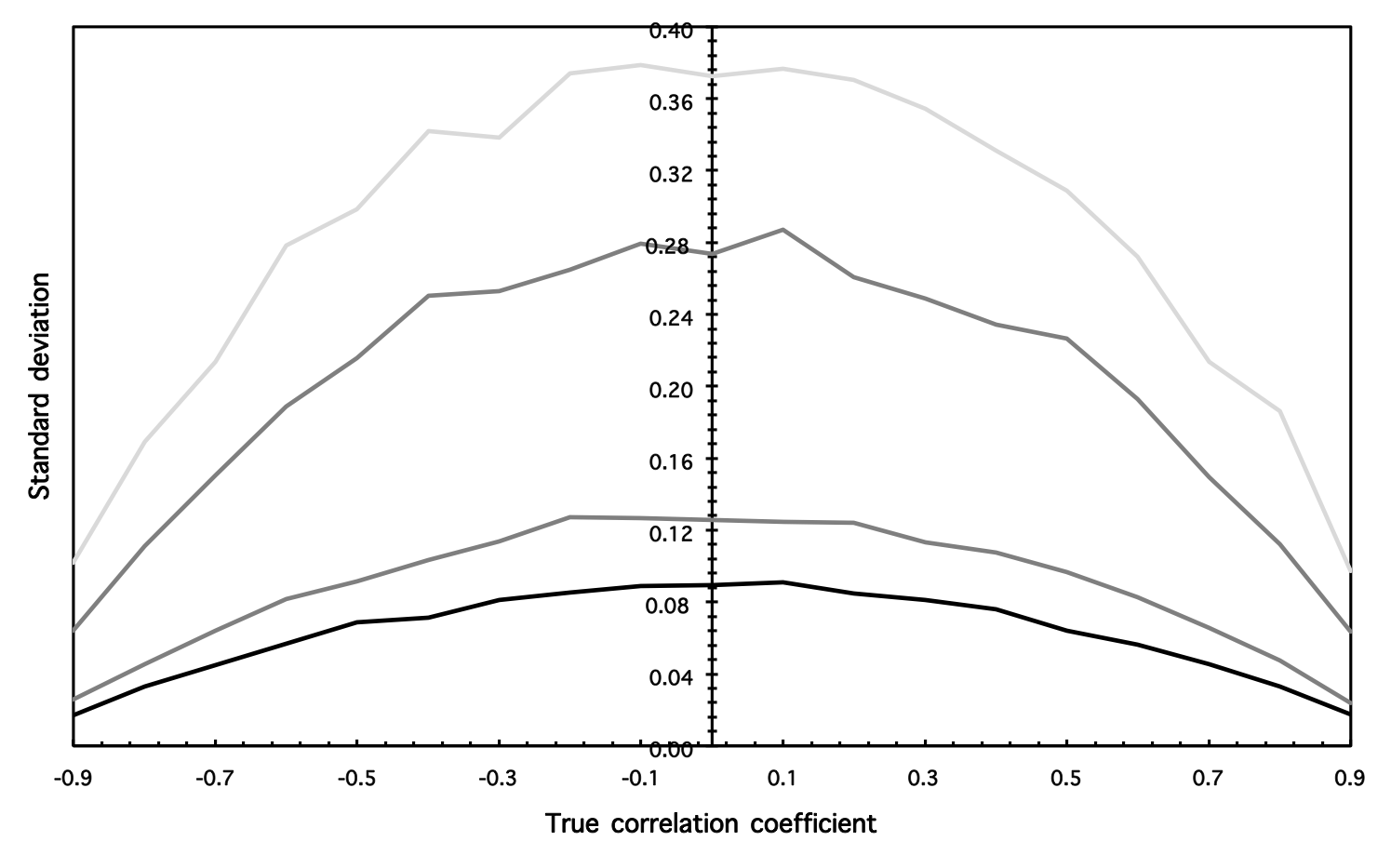}&\includegraphics[width=45mm]{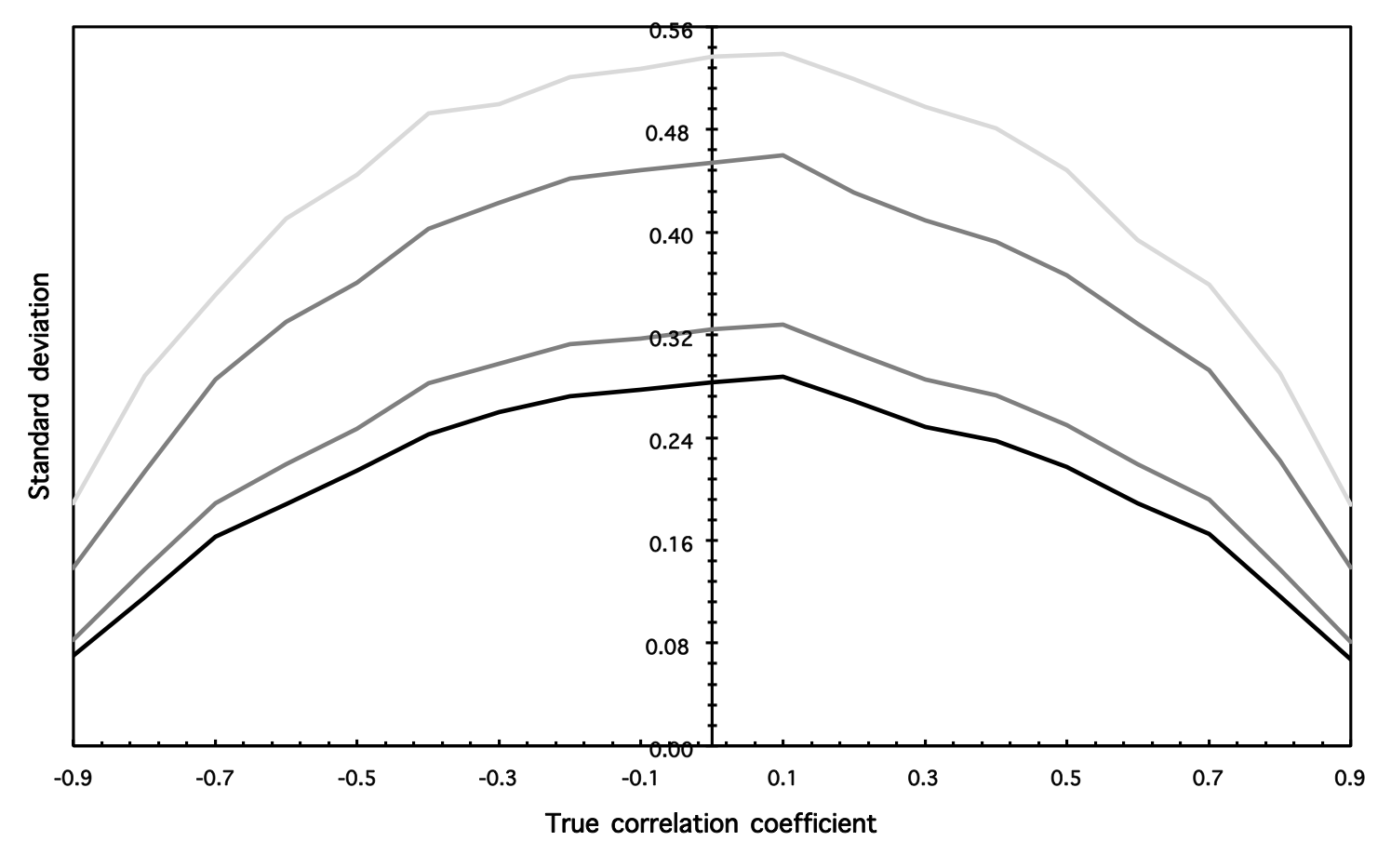}\\
\end{tabular}
\caption{\textbf{Standard deviations of DCCA correlation coefficients for different fractional integration parameters $d$ II.} \footnotesize{Results for the time series of length $T=5000$ are shown here. Notation of Fig. \ref{fig2} is used.}\label{fig4}}
\end{center}
\end{figure}

Figs. \ref{fig1} and \ref{fig3} show the median values and the 95\% confidence intervals of the DCCA coefficients for $T=1000$ and $T=5000$, respectively, for various scales $s$, correlation coefficients $\rho_{\varepsilon\nu}$ and the fractional integration levels $d$. The results are based on 1000 simulations of Eq. \ref{eq:ARFIMA1} with specific parameters setting. In Figs. \ref{fig2} and \ref{fig4}, the standard deviations of the estimated DCCA coefficients are illustrated as additional measures of the DCCA coefficient precision.

The results share several common points. Firstly, the DCCA coefficient is an unbiased estimator of the true correlation coefficient regardless all the parameters settings we apply. Secondly, the estimates are more precise for higher absolute values of the true correlation coefficient. Thirdly, the precision (measured by the standard deviations) is approximately symmetric around the zero correlation. Fourthly, the estimates are more precise for the lower scales. Fifthly, the precision of the estimates decreases with increasing $d$. And sixthly, the precision of the estimates does not vary much with increasing time series length; the dominant parameter in this case seems to be the scale $s$ and its connection to the time series length $T$ rather than $T$ alone. The most important of these points is the fact that the DCCA coefficient is able to estimate the true correlation coefficient with no bias even for strongly non-stationary series. Even though the confidence intervals widen markedly between $d=1.1$ and $d=1.4$, the DCCA coefficient remains unbiased. By this analysis, we strongly support the claims of Zebende \cite{Zebende2011} and Podobnik \textit{et al.} \cite{Podobnik2011} that the DCCA coefficient can be used  to measure correlation between non-stationary time series.  

\section{Discussion and conclusions}

\begin{figure}[!htbp]
\begin{center}
\begin{tabular}{ccc}
\includegraphics[width=45mm]{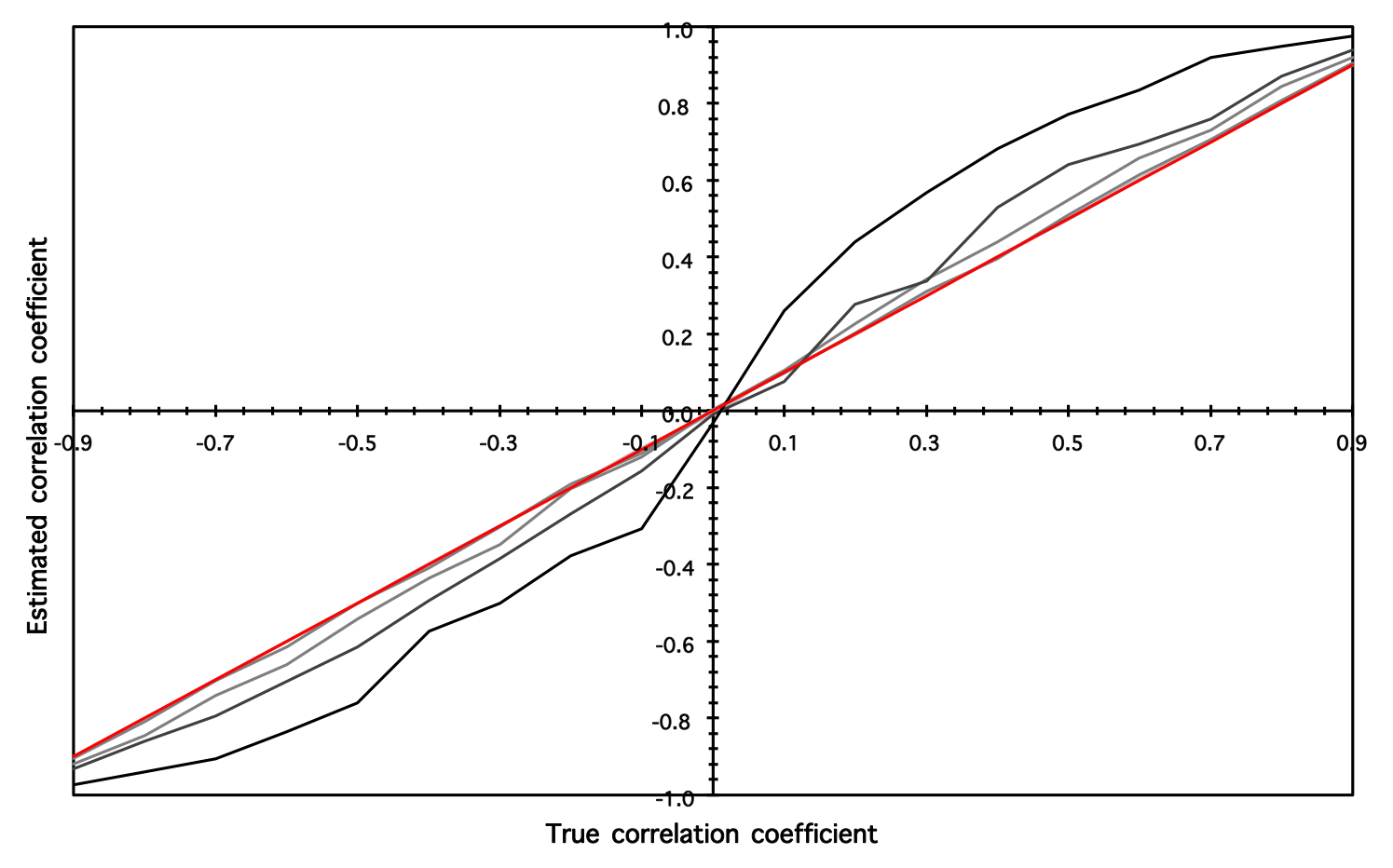}&\includegraphics[width=45mm]{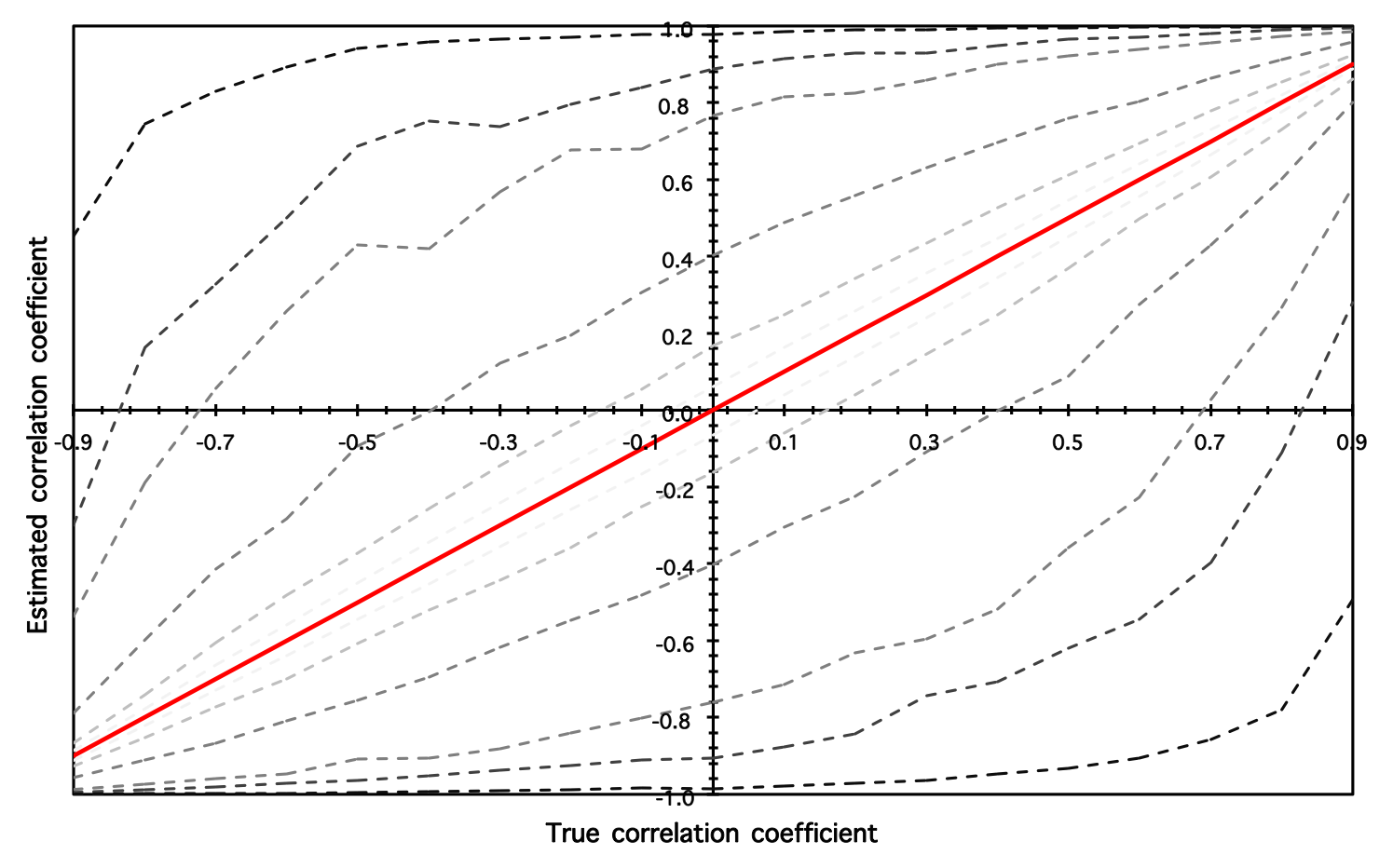}&\includegraphics[width=45mm]{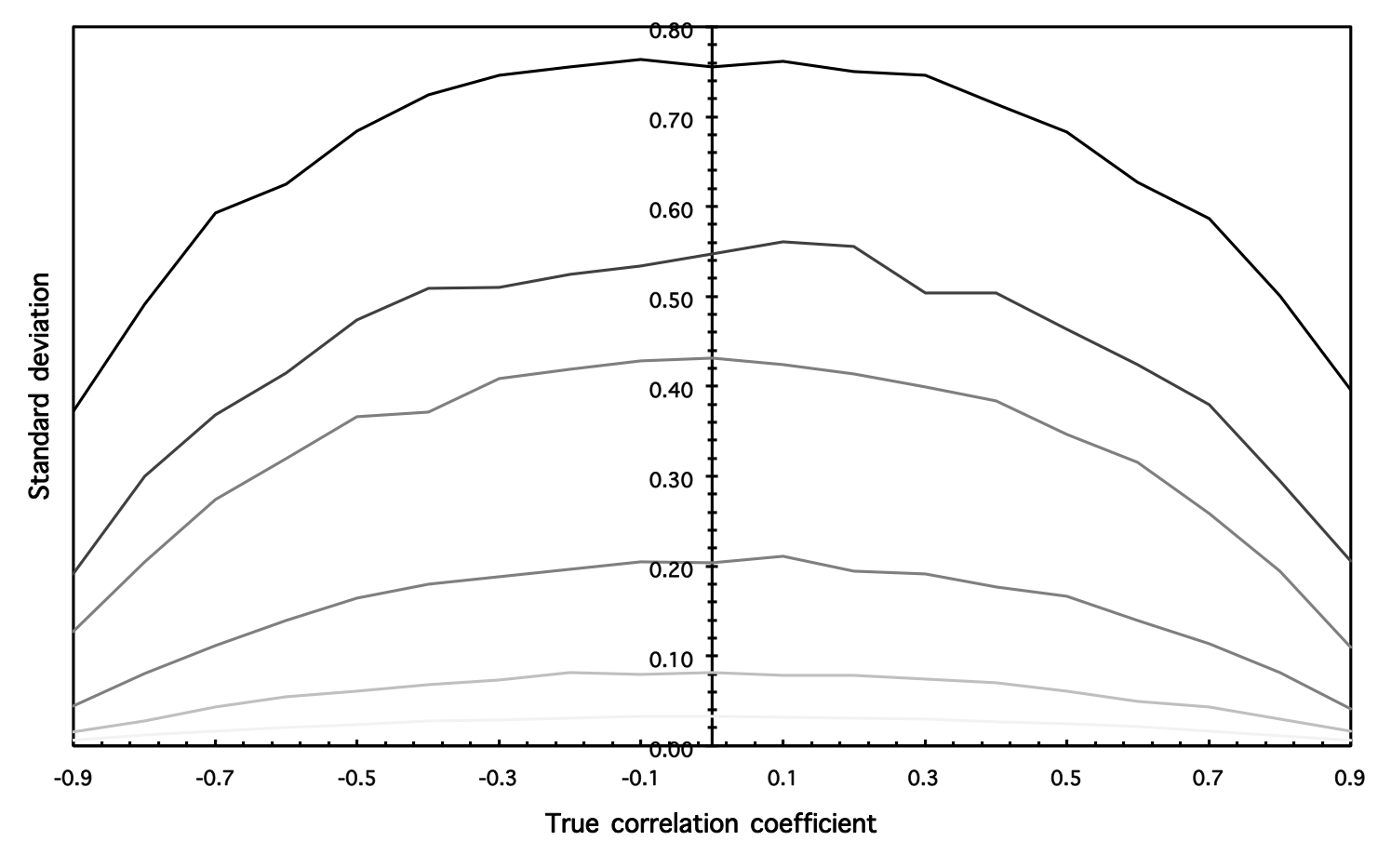}\\
\includegraphics[width=45mm]{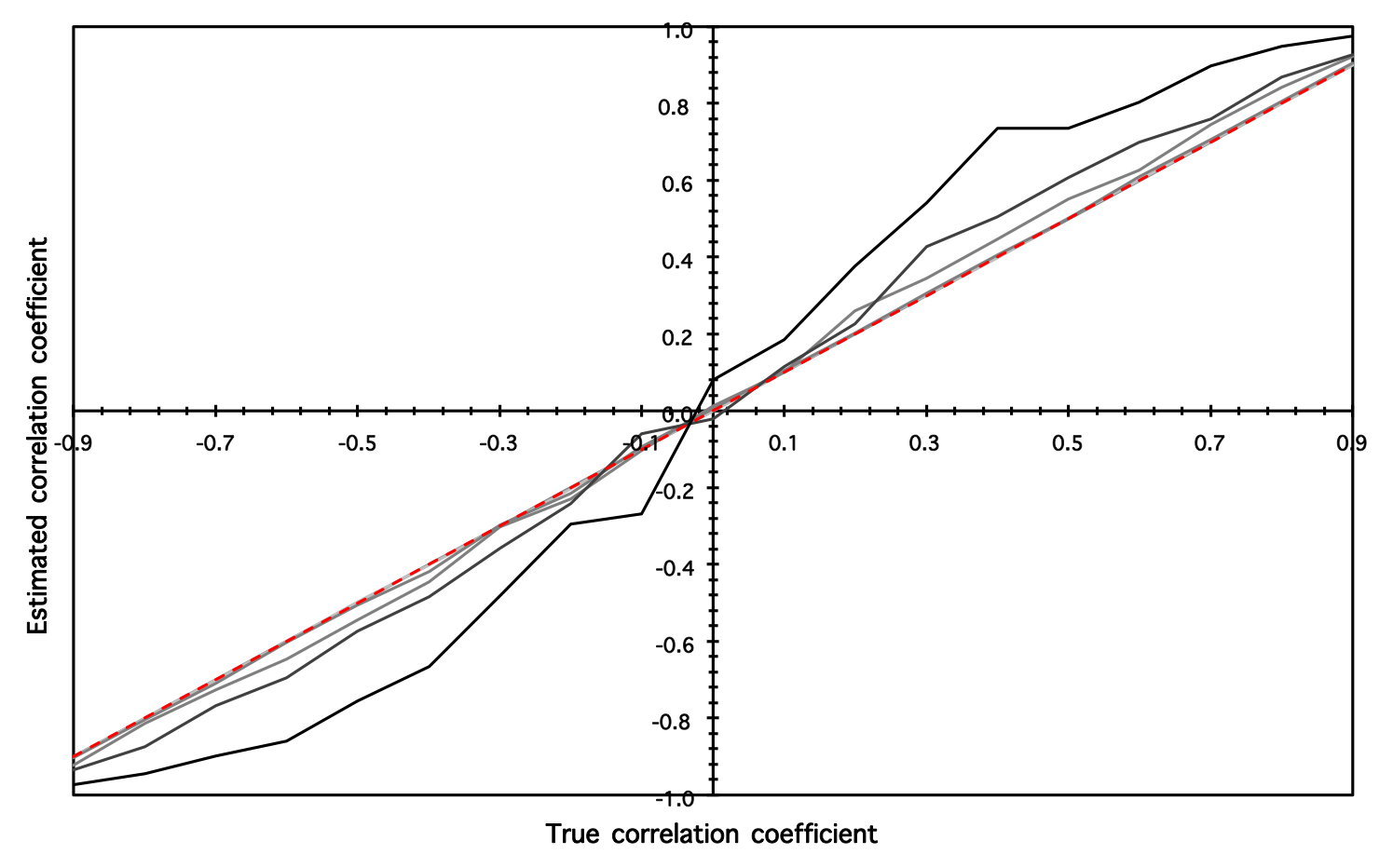}&\includegraphics[width=45mm]{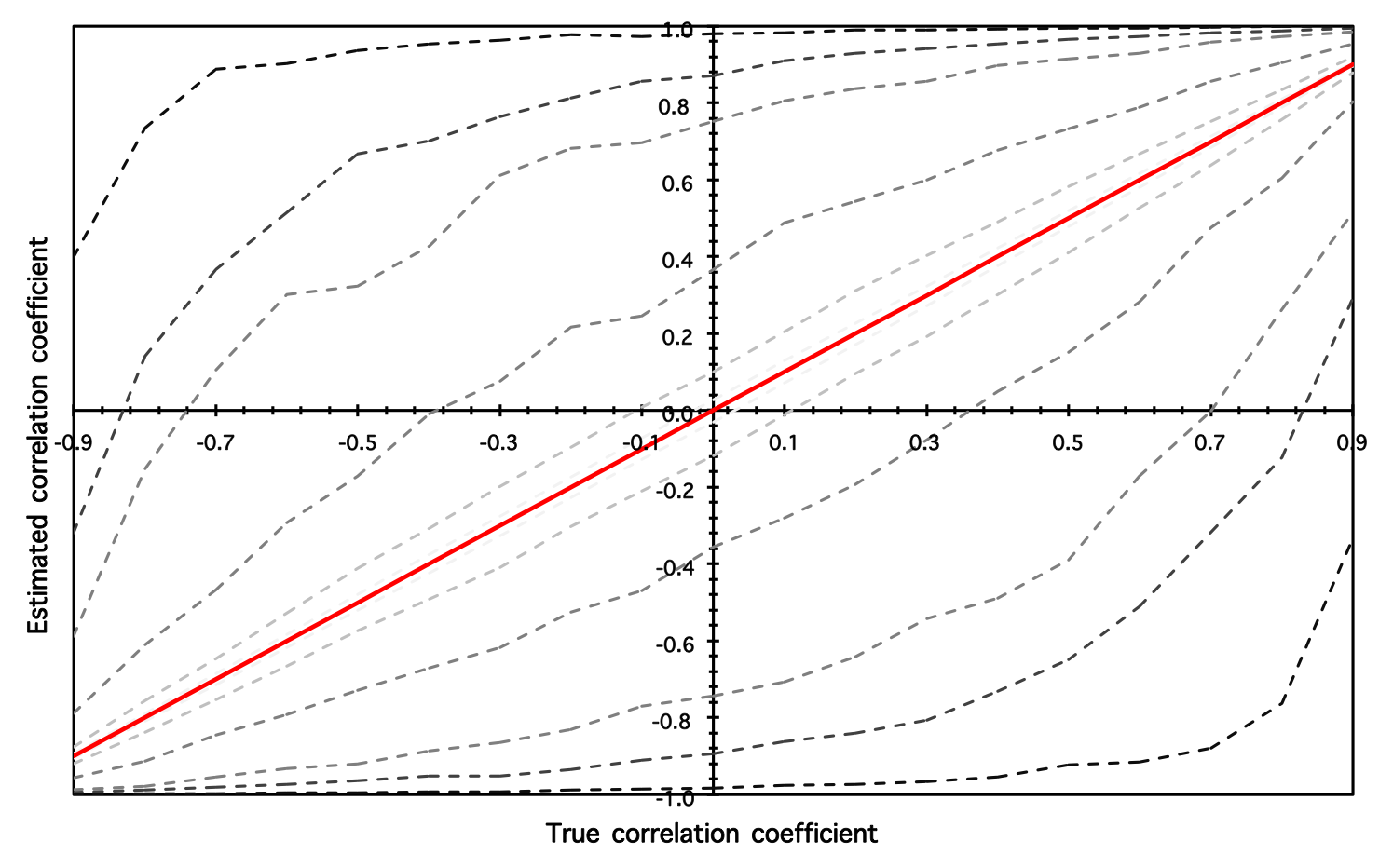}&\includegraphics[width=45mm]{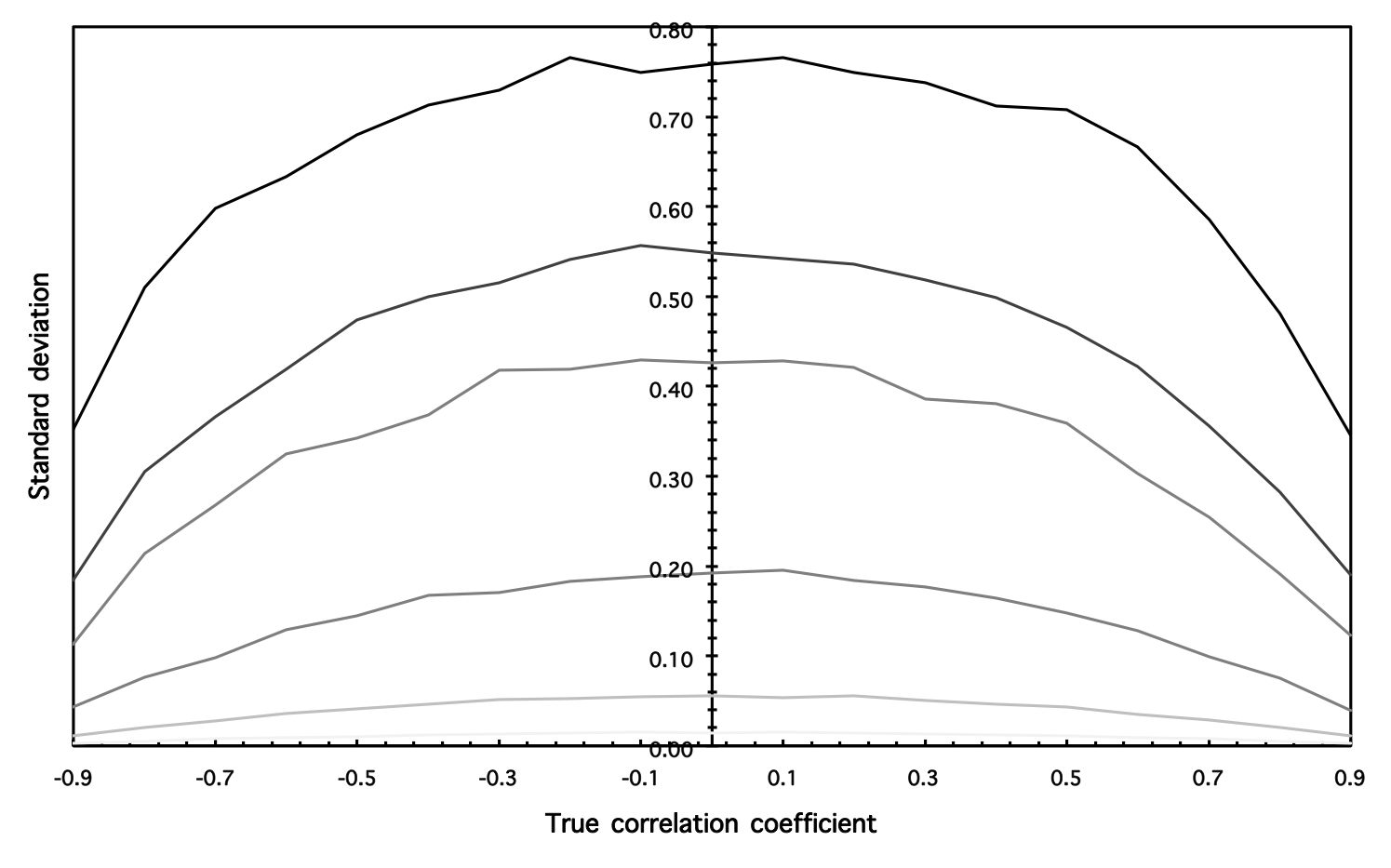}\\
\end{tabular}
\caption{\textbf{Pearson's correlation coefficients for different fractional integration parameters $d$.} \footnotesize{Results for both $T=1000$ (top) and $T=5000$ (bottom) are shown. Different shades of grey are used here to distinguish between values of $d$ ranging from $d=0.1$ (the lightest shade) to $d=1.4$ (the darkest shade). The figures on the left show the median values of 1000 simulations for the given setting. In the middle, the 95\% confidence intervals are represented by the dashed lines. The red lines represent the true values of correlations. The right panel shows the standard deviations of the simulations for given settings.}\label{fig5}}
\end{center}
\end{figure}

In order to support our results and also to stress a need for a precise estimator of the correlation coefficient between non-stationary series, we present the results of simulations of the standardly used Pearson's correlation coefficient in the same simulation setting. As the simulations now lose scale parameter $s$, we can present the results more easily in Fig. \ref{fig5}. In the figure, we again observe several regularities. Firstly, the estimator becomes biased for non-zero correlations and the bias increases (in absolute value) with increasing strength of non-stationarity (increasing $d$). Secondly, the standard deviation increases markedly with the parameter $d$. And thirdly, the confidence intervals of the Pearson's correlation coefficient become extremely wide for non-stationary series. By looking at Fig. \ref{fig5}, we observe that even for a rather weak non-stationarity of $d=0.6$ and zero true correlation, the confidence intervals range between approximately -0.4 and 0.4 (compared to approximately -0.13 and 0.13 for the DCCA coefficient with $s=T/100$). For the non-stationary slightly below the unit-root -- $d=0.9$ -- the range widens to an interval between -0.75 and 0.75 (compared to approximately -0.16 and 0.16 for the DCCA coefficient with $s=T/100$). For the stronger forms of non-stationarity, the confidence intervals cover almost the whole range of the correlation coefficients. This again does not change much with an increasing time series length. These results indicate that the standard Pearson's coefficient is practically useless for non-stationary time series.

Contrary to the results for the Pearson's correlation coefficient, we have shown that the DCCA coefficient is able to estimate the true correlation coefficient between series precisely regardless the non-stationarity strength. Even though the performance varies with some of the parameters, the coefficient remains a very promising tool for measuring the dependence between non-stationary series.

\section*{Acknowledgements}
The support from the Grant Agency of the Czech Republic (GACR) under projects P402/11/0948 and 402/09/0965, and project SVV 267 504 is gratefully acknowledged.

\bibliography{DCCA}
\bibliographystyle{chicago}

\end{document}